\documentclass[conference, 10pt]{IEEEtran}

\usepackage[textsize=tiny]{todonotes}
\usepackage{nicefrac}
\usepackage{csquotes}
\usepackage{mathtools,amssymb,amsthm,amsmath}
\usepackage{booktabs}
\usepackage[noadjust]{cite}
\usepackage{enumitem}
\usepackage{siunitx}
\usepackage{url}
\usepackage{graphicx}
\usepackage{setspace}
\usepackage{comment}
\usepackage[capitalise]{cleveref}
\usepackage{blkarray}
\usepackage{bm}
\usepackage{subcaption}
\usepackage{soul}
\usepackage{listings}
\usepackage{textgreek}
\usepackage{diagbox}
\usepackage{multirow}
\usepackage{xcolor}
\usepackage{pgfplots}
\usepackage{tikz}
\usepackage{float}

\usetikzlibrary{backgrounds,fit,decorations.pathreplacing,calc,matrix,arrows.meta}

\renewcommand{\baselinestretch}{.98}

\renewcommand\IEEEkeywordsname{Keywords}

\begin{document}

\title{\Large\bf Accelerated Computational Fluid Dynamics Simulations of \\ Microfluidic Devices by Exploiting Higher Levels of Abstraction}

\date{\today}
\author{\IEEEauthorblockN{Michel Takken\IEEEauthorrefmark{1} \qquad \qquad \qquad 
Robert Wille\IEEEauthorrefmark{1}\IEEEauthorrefmark{2}}
\IEEEauthorblockA{\IEEEauthorrefmark{1}School of Computation, Information and Technology, Technical University of Munich, Germany\\
\IEEEauthorrefmark{2}Software Competence Center Hagenberg GmbH (SCCH), Austria\\
michel.takken@tum.de \qquad \quad \qquad robert.wille@tum.de\\
\url{https://www.cda.cit.tum.de/research/microfluidics/}}}

\maketitle

\begin{abstract}
The design of microfluidic devices is a cumbersome and tedious process that can be significantly improved by simulation. Methods based on \emph{{Computational Fluid Dynamics}}~(CFD) are considered state-of-the-art, but require extensive compute time---oftentimes limiting the size of microfluidic devices that can be simulated. Simulation methods that abstract the underlying physics on a higher level generally provide results instantly, but the fidelity of these methods is usually worse. In this work, a simulation method that accelerates CFD simulations by exploiting simulation methods on higher levels of abstraction is proposed. Case studies confirm that the proposed method accelerates CFD simulations by multiple factors (often several orders of magnitude) while maintaining the fidelity of CFD simulations.
\end{abstract}

\begin{IEEEkeywords}
    lab-on-a-chip; microfluidics; simulation; CFD.
\end{IEEEkeywords}

\section{Introduction}
\label{sec:intro}
A microfluidic device, or a \emph{{Lab-on-a-Chip}} 
(LoC), is a device that performs lab operations on the microscale through a set of fluid manipulations \cite{Convery2019}. Such devices are commonly used for, e.g., personalized medical care \cite{Mathur2020}, point-of-care diagnostics~\cite{Pandey2018} (well-known examples of such devices are pregnancy tests \cite{Schonhorn2014} or the SARS-CoV-2 tests \cite{Yang2021}) and the food industry~\cite{He2020}. They have recently been proven to be more widely applicable in, e.g., geosciences \cite{Zhu2022} or fuel cell technology \cite{Wang2021FC}. In that regard, much potential lies in microfluidic devices.

The development of those microfluidic devices is a cumbersome and tedious process that often requires multiple expensive and \mbox{time-consuming} design cycles \cite{Grimmer2018}. To advance the design of microfluidic devices, reliable and quick simulation methods are necessary to predict whether a design works as intended. The simulation of Newtonian flow through a single microchannel can be performed using the \mbox{Hagen--Poiseuille Law}~\cite{Kirby2010} and the flow profile across the channel can be assumed to be parabolic \cite{Bruus2007}. However, channel-based microfluidic devices consist of multiple interconnected microchannels, which require more extensive simulation methods.

To this end, most designers utilize methods from \emph{{Computational Fluid Dynamics}} (CFD), such as the \emph{{Finite Volume Method}} (FVM, \cite{Ferziger2019}), \emph{{Finite Element Method}} (FEM, \cite{Blazek2015}), or the \emph{{Lattice Boltzmann Method}} (LBM, \cite{Kruger2017}), to obtain results of good fidelity. In recent overviews for microfluidics modeling, the FVM \cite{Rapp2022}, FEM \cite{Rapp2022}, and LBM \cite{Bazaz2020} are listed as numerical approaches to solving the \emph{{Navier--Stokes Equations}} (NSE) for microfluidics. Another recent work describes the workflow of setting up simulations for microfluidic devices~\cite{Ebner2023} with {OpenFOAM v9.0} \cite{Jasak2009}, {which uses the FVM}
. Hence, these simulation methods can be considered \mbox{state-of-the-art} for modeling microfluidic devices.

However, CFD simulations can take up to days or even weeks, even on dedicated workstations \cite{Glatzel2008}. In practice, this obviously limits the use of CFD simulations for microfluidic devices to single components of the device~\cite{Takken2022}. To fully understand and design a harmonious device, it is critical to simulate the behavior of \emph{{all}} 
components \emph{{and}} 
their interaction with each other.

Alternatively, high abstraction simulation methods, i.e., simulation methods that abstract the underlying physics on a higher level (also known as reduced-order modeling), generally simulate microfluidic devices almost instantly (i.e., in less than a second)~\cite{Takken2022}. An example of such a method is to draw an analogy between the Hagen--Poiseuille {law} 
and Ohm's law and apply analogous methods from electrical circuit engineering to \mbox{{channel-based}} 
microfluidic devices \cite{Oh2012}. This approach is not limited to basic fluid flow, but can also be applied to problems with, e.g., droplets~\cite{Grimmer2019}, or \mbox{capillary-driven} flows for paper-based microfluidics \cite{Elizalde2015}. Such methods may not provide results of comparable fidelity, but can still simulate some parts of microfluidic devices, e.g., channels, with relatively good accuracy~\cite{Takken2022}.

In this work, the nature of these high abstraction simulation methods is exploited to substantially accelerate CFD simulations of microfluidic devices with hardly any loss of fidelity. To this end, a two-stage approach is proposed: First, regions of the microfluidic device are identified that can sufficiently be simulated at a high level of abstraction. Afterward, the corresponding simulations are conducted, and the respectively obtained results are communicated between the simulation engines.
Case studies (using continuous \mbox{channel-based} microfluidic devices as a representative) confirm the potential of this approach: The proposed approach does not only generate simulation results faster than the original CFD method, but constantly does so by several factors or even several orders of magnitude---while, at the same time, maintaining the fidelity of the results.

The remainder of this work is structured as follows: First, we review the simulation methods for microfluidic devices in more detail, focusing on methods on low and high abstraction levels. Afterward, the accelerated CFD simulation method is described in a general fashion in \cref{sec:GeneralScheme}. Implementation details of this method are then provided in \cref{sec:Application}. Finally, we demonstrate the resulting solution and compare it against solutions of CFD simulations for a set of test cases in \cref{sec:Results}---confirming that the accelerated method is faster than the CFD method for all test cases while maintaining the fidelity. Finally, the paper is concluded in \cref{sec:Conclusion}.

\section{Simulation Methods for Microfluidic Devices}
\label{sec:background}

Simulation methods for microfluidic devices can be categorized into methods based on \emph{{Computational Fluid Dynamics}}~(CFD, \mbox{\cite{Ferziger2019,Blazek2015,Kruger2017}}) on the low abstraction level, and methods on high abstraction levels~\cite{Oh2012,Grimmer2019,Elizalde2015}, sometimes referred to as 1D methods. CFD methods can be considered state-of-the-art in the design of microfluidic devices \cite{Rapp2022,Bazaz2020,Ebner2023}, whereas the high abstraction level methods are often used to derive initial estimates during the design process \cite{Grimmer2018}. In this section, we review the CFD and high abstraction level simulation methods.

\subsection{Review of CFD Methods}
A low abstraction simulation of microfluidic devices can be obtained through methods from \textit{{Computational Fluid Dynamics}}~\mbox{(CFD, \cite{Ferziger2019,Blazek2015,Kruger2017}}). In CFD, the fluid behavior is modeled by the \emph{\mbox{{Navier--Stokes}} {Equations}}~\mbox{(NSE)}, which are the fundamental governing equations for fluid dynamics. We restrict ourselves to the incompressible NSE \cite{Ferziger2019,Blazek2015,Kruger2017}, which are given by the mass equation

\begin{equation}\label{eq:NSE-mass}
	\nabla \cdot \bm{u} = 0,
\end{equation}

\noindent where \(\bm{u}\) is the flow velocity vector, and the momentum equation

\begin{equation}\label{eq:NSE-momentum}
	\frac{\partial}{\partial t} \left(\bm{u}\right) + \nabla \cdot \left( \bm{u} \bm{u}^T \right) = -\frac{1}{\rho} \nabla p + \nu \nabla^2 \bm{u},
\end{equation}

\noindent where \(\rho\) is the fluid density, \(p\) is the pressure, and \(\nu\) is the kinematic viscosity.

The analytical solution of the incompressible NSE exists only for a few simple fluid dynamics problems. For microfluidic devices, the equations generally have to be solved numerically. To this end, a wide variety of numerical methods has been developed. For example: 

\begin{itemize}
    \item The \textit{{Finite Volume Method}}~\mbox{(FVM, \cite{Ferziger2019})}. The FVM splits the computational domain into grid cells, and the NSE are solved numerically on each grid cell. By this, averaged values for the flow velocity and pressure are obtained in each cell.
    
    \item The \textit{{Finite Element Method}}~\mbox{(FEM, \cite{Blazek2015})}. Similar to the FVM, the domain is split into cells. However, the solution is represented by a set of elements, e.g., polynomials. 
    
    \item The \textit{{Lattice Boltzmann Method}}~\mbox{(LBM, \cite{Kruger2017})}. Rather than solving the NSE, this method solves the Boltzmann equation---which can be proven, through the \mbox{Chapman--Enskog} 
     theory, to macroscopically solve the NSE \cite{Kruger2017}.
\end{itemize}

CFD methods have in common that they, in a general sense, acquire results of good fidelity and are, therefore, helpful for the design of delicate microfluidic devices or components. However, CFD methods require significant computational resources---in terms of computational memory and time. A CFD simulation of entire microfluidic devices can be hard to get right, and compute times can get up to days or weeks, even on dedicated workstations \cite{Glatzel2008}.

\subsection{Review of Methods on High Abstraction Levels}
\label{sec:Review-high-abstraction}

Methods with a high level of abstraction simulate microfluidic devices by abstracting the underlying physics on a high level. Generally, these methods are based on solutions that can be obtained analytically for fairly simple problems. They map the results to similar problems under a set of simplifying assumptions, or models are obtained empirically through fitted data from experiments or \mbox{pre-simulations} \cite{Fink2021}. These methods are generally of poor fidelity, but results for entire microfluidic devices are usually acquired instantly (i.e., in less than a second) \cite{Takken2022}. More precisely, high abstraction simulation methods have been proposed for the following microfluidic platforms \cite{Tsur2020}:

\begin{itemize}

    \item \emph{{Continuous channel-based microfluidics}}. This platform consists of a network of rectangular channels with width and height in the order of micrometers. Liquid flow through these channels can practically always be assumed to be laminar, and the flow profile in a channel can be accurately solved using the \mbox{Hagen--Poiseuille} law, i.e.,
    \begin{equation}\label{eq:Hagen-Poiseuille}
        \Delta p = Q \cdot R_H,
    \end{equation}
    where \(Q\) is the flow rate and \(R_H\) is the hydraulic 
    resistance of a channel. Using the \textit{{Modified Nodal Analysis}}~(\mbox{MNA}, \cite{Oh2012}), the pressure and flow rates of all channels in a connected network can be calculated.
    
    \item \emph{{Droplet-based microfluidics}}. This platform has a network of channels, similar to the continuous \mbox{channel-based} microfluidics platform, with additional droplets of a fluid that is immiscible with the carrier fluid (continuous phase). The hydraulic resistance in \cref{eq:Hagen-Poiseuille} can be split into the resistance of the channel $R_H^\text{channel}$ and the resistance of a droplet $R_H^\text{droplet}$ present in that channel, i.e.,
    \begin{equation}
        R_H = R_H^\text{channel} + R_H^\text{droplet}.
    \end{equation}

    Based on this and the MNA, this platform can be simulated on a high abstraction level~\cite{Grimmer2019}.
    
    \item \emph{{Paper-based microfluidics}}. In paper-based microfluidics, a liquid is transported through a two-dimensional sheet of paper using capillary force. The \mbox{one-dimensional} transport of a liquid front through a porous medium is given by the Washburn equation, i.e.,
    \begin{equation}\label{eq:Washburn}
        L^2 = \frac{\gamma D t}{4 \mu},
    \end{equation}
    where \(L\) is the traversed distance of the fluid front, \(\gamma\) is the effective surface tension, \(D\) is the diffusivity coefficient, \(t\) is time, and \(\mu\) is the dynamic viscosity. 
    Based on \cref{eq:Washburn}, the capillary transportation of fluid can be simulated for porous channels with arbitrary \mbox{cross-sectional} shapes \cite{Elizalde2015}.
    
\end{itemize}

\section{Accelerating CFD Simulations}
\label{sec:GeneralScheme}

Motivated by the fast computation time of methods with a high level of abstraction, the possibility to accelerate CFD simulations by exploiting said methods is investigated. In this work, we aim to accelerate CFD simulations for steady-state flow of continuous channel-based microfluidic devices as a representative. Firstly, the semantics of continuous \mbox{channel-based} microfluidics are covered in this section, and the potential for a faster simulation is highlighted. Afterward, a method will be proposed that explicitly exploits that potential and accelerates CFD simulations. Based on that, implementation details for this method and a summary of corresponding case studies, including evaluation results, are provided in the following sections.

\subsection{Continuous Channel-Based Microfluidics}

\begin{figure}[H]
\centering
\includegraphics[width=0.8\columnwidth]{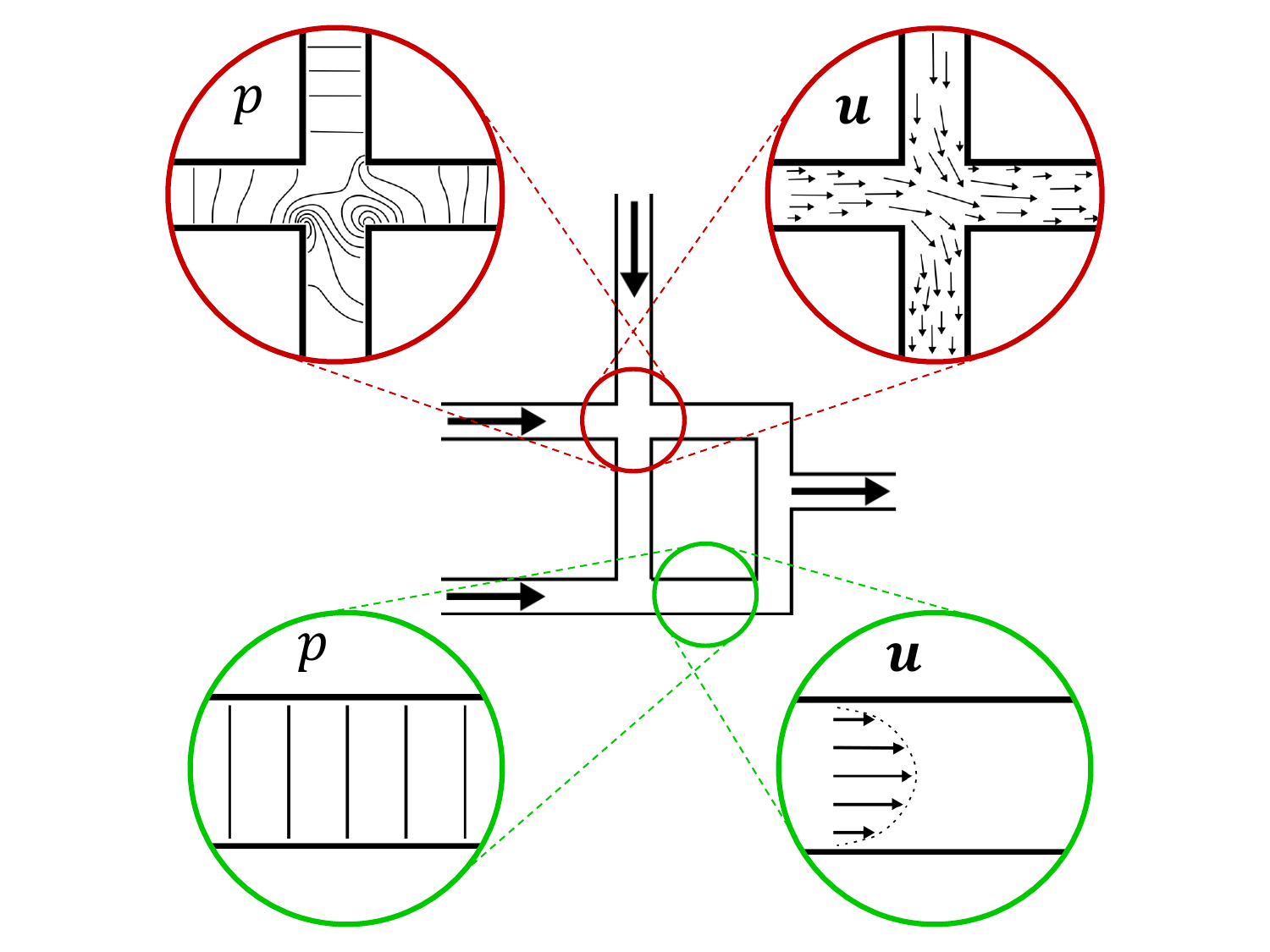}
\caption{Example network for continuous channel-based microfluidics. The pressure~$p$ and velocity fields~$\bm{u}$ are shown in detail for a crossing (red) and a straight channel section (green).}
\label{fig:ExampleNetwork}
\end{figure}

In continuous channel-based microfluidics, we consider a network as sketched in the middle of Figure \ref{fig:ExampleNetwork}. This example network has three inlets and one outlet and contains a homogeneous fluid (no mixture). It is depicted here as a \mbox{two-dimensional} network of channels with width $w$, and the extension to the third dimension for real-world microfluidic devices can be performed by adding a height parameter $h$. Furthermore, without loss of generality, we assume an adiabatic system and ignore gravity effects, such that the only relevant fields are the pressure~$p$ and velocity~\(\bm{u}\) of the fluid. 

However, as in this setup, the pressure and velocity fields of a fluid in a microfluidic network are generally complex and require dedicated methods to solve \cref{eq:NSE-mass,eq:NSE-momentum}. This is sketched in \cref{fig:ExampleNetwork} for the pressure and velocity field of the \mbox{steady-state} flow at the location where channels cross (the red circles at the top). The contour lines of the pressure field and vectors of the velocity field are chaotic and not easy to predict. On the other hand, if we look at the pressure and velocity fields of a straight channel section (the green circles at the bottom), we notice a more organized and streamlined flow. The contour lines of the pressure field are straight, evenly spaced, and perpendicular to the channel. We can see here that, for straight channels, the pressure is a \mbox{point-value} along the channel and equal over the channel's \mbox{cross-section}. The velocity vectors are also organized and show a parabolic (for Newtonian fluids~\cite{Takken2022}) flow profile, as can be found, for a two-dimensional channel, by using
     
\begin{equation}
    \bm{u}(x,y) = \left(\frac{y(h-y)}{2\mu}\frac{\partial p}{\partial x} \;\;\;\;\;\; 0\right)^T,
\end{equation}

\noindent where $x$ is the direction parallel to the channel, and $y$ is perpendicular to the channel. This is the cornerstone, on which \cref{eq:Hagen-Poiseuille} and, therefore, the high abstraction simulation method for continuous channel-based microfluidics is based.

\subsection{The Proposed Accelerated Method}

Based on the observation above, we can exploit the high abstraction level simulation method in regions where the flow is highly organized (green circles in \cref{fig:ExampleNetwork}) and use CFD simulations for regions where the flow is chaotic and hard to predict (red circles in \cref{fig:ExampleNetwork}). Exploiting the high-speed simulation feature of methods on a high abstraction level not only significantly reduces the compute time for large microfluidic devices (resulting in more favorable scaling of simulations) but also reduces the required memory. 
To this end, two steps must be taken to apply the accelerated method:
\begin{enumerate}
    \item The required fidelity for the {complete network} $\Omega$ must be defined and $\Omega$ must be split into $\Omega_\text{low}${-} and \mbox{$\Omega_\text{high}${-}{regions}}
    , such that \mbox{\(\Omega_\text{high} \cup \Omega_\text{low} = \Omega\)} and \(\Omega_\text{high}\) is as large as possible. Here, $\Omega_\text{low}$ is the set of regions that require a good fidelity method and should be simulated on a low abstraction level, whereas $\Omega_\text{high}$ is the set of regions that can be simulated relatively accurately using methods on a high abstraction level. 
    \item The resulting pressure and velocity fields of the corresponding simulation methods must be equal (or at least in close vicinity) on the {interface} \(\Gamma = \Omega_\text{high} \, \cap \, \Omega_\text{low}\) to ensure continuity of the {complete solution} \textphi{} in the complete network $\Omega$. This means that the pressure and velocity values must be communicated between the simulation methods and subsequently updated in $\Omega_\text{low}$ and~$\Omega_\text{high}$.
\end{enumerate}

In the next section, the implementation details on both steps are illustrated.

\section{Implementation Details}
\label{sec:Application}
To properly describe the implementation details for the method proposed above, the continuous channel-based microfluidic device sketched in \cref{fig:ExampleNetwork} is used as a running example. Furthermore, the LBM and MNA (as reviewed in \cref{sec:background}) are used as simulation methods for $\Omega_\text{low}$ and $\Omega_\text{high}$, respectively. The idea proposed above can be realized as follows.

\subsection{Step 1: Identifying the Required Fidelity}

\begin{figure}[H]
	\begin{subfigure}[b]{0.45\columnwidth}
		\centering
		\includegraphics[width=0.8\textwidth]{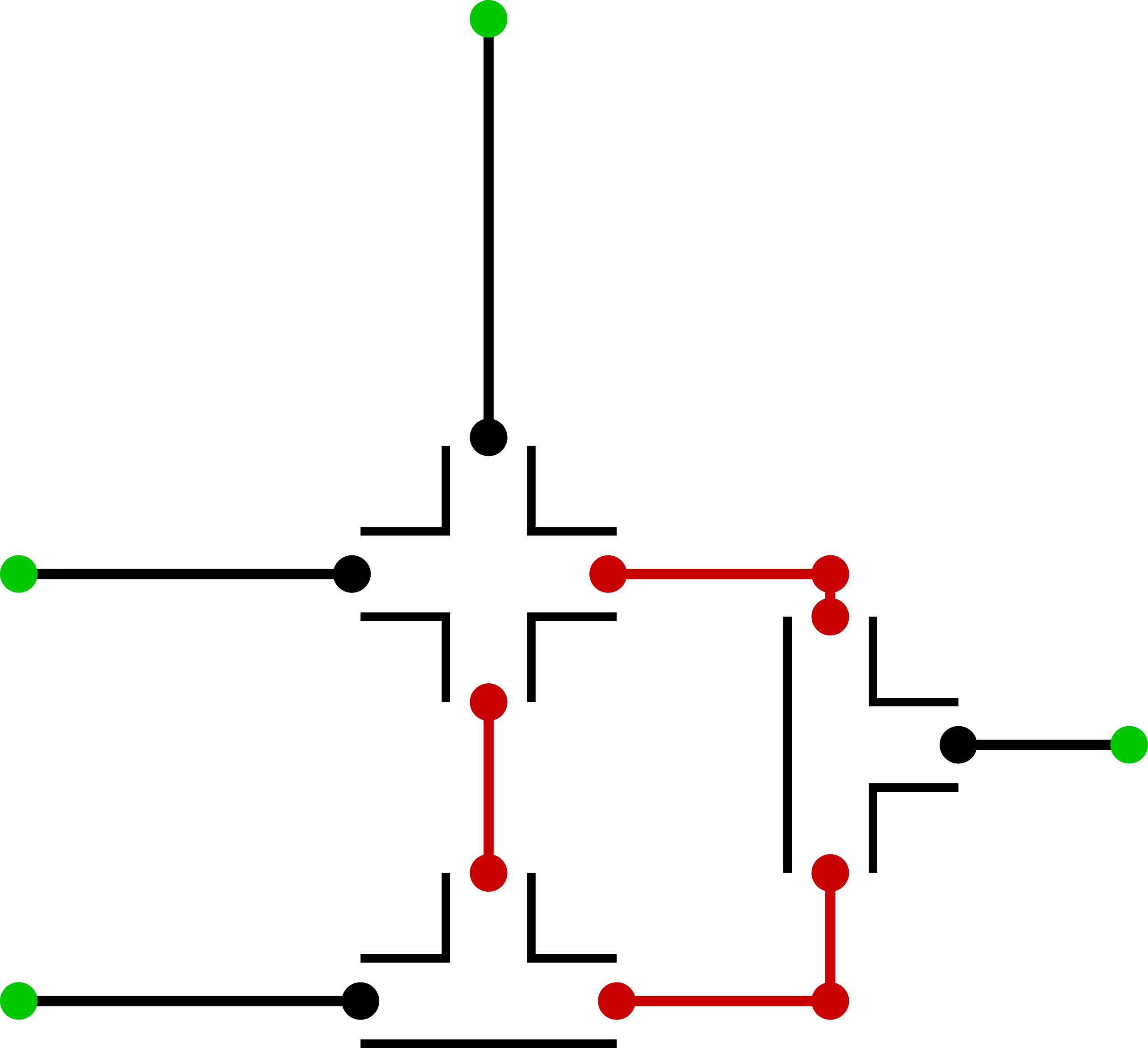}
\captionsetup{justification=centering}
		\caption{ }
		\label{subfig:ExampleSeparated}
	\end{subfigure}
	\hfill
	\begin{subfigure}[b]{0.45\columnwidth}
		\centering
		\includegraphics[width=0.8\textwidth]{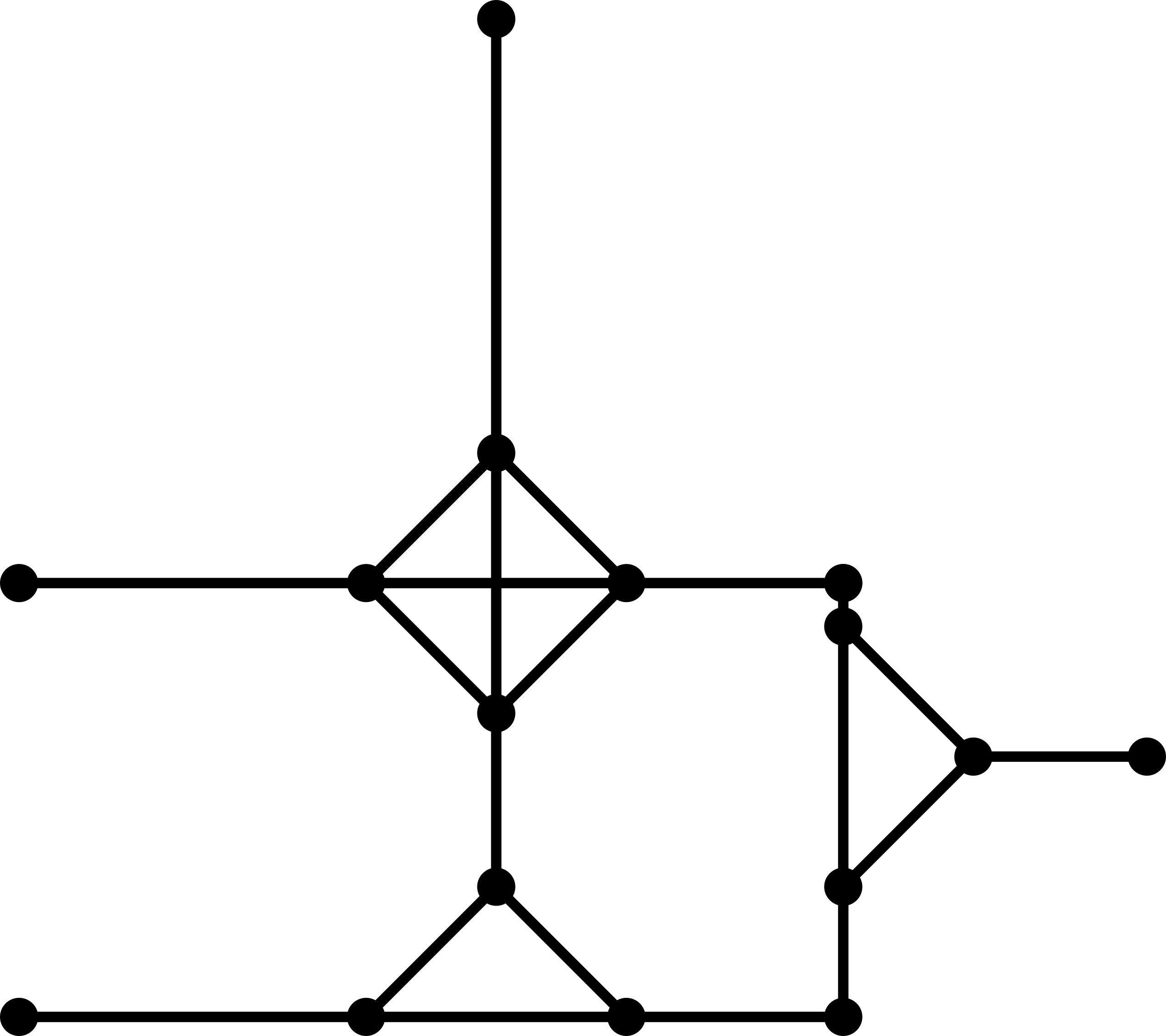}
\captionsetup{justification=centering}
		\caption{ }
		\label{subfig:ExampleInitial}
	\end{subfigure}
	\caption{{The} 
 {example} 
 network in the proposed method. {(\textbf{a})} 
 The separation of the example network into $\Omega_\text{low}$ and $\Omega_\text{high}$. (\textbf{b}) The network during the initial iteration; $\Omega_\text{low}$ is replaced by fully connected graphs, and the resulting network is used to find $q^0$.}
\label{fig:Figure2}
\end{figure}

Recall that the first step aims at identifying (ideally many) $\Omega_\text{high}$-regions where a high abstraction simulation method is sufficient and, hence, can be utilized to accelerate the required computations. 
As sketched before in \cref{fig:ExampleNetwork}, those regions can usually be identified easily. For example, straight channels belong to~$\Omega_\text{high}$, while, e.g., crossings and \mbox{T-junctions} better remain in~$\Omega_\text{low}$. \textGamma{} should be located inside a straight channel, sufficiently far from, e.g., a crossing or junction, where the flow is organized and the pressure can be regarded as a point-value along the channel. The resulting separation of the network is illustrated in {Figure} 
 \ref{fig:Figure2}a, where the two-dimensional regions represent $\Omega_\text{low}$, and the lines and nodes constitute $\Omega_\text{high}$. 

Please note that the identification step is not restricted to straight channels, crossings and T-junctions. Also, arbitrary channel shapes or components on the microfluidic device (e.g., heaters, mixers, droplet generators, etc.) could be identified as $\Omega_\text{low}$ as long as corresponding high abstraction level simulation methods are available.

\subsection{Step 2: Defining the Communication}

The second step aims at ensuring that the complete solution~\textphi{} is continuous in $\Omega$. Hence, the correspondingly obtained values (here, pressure and velocity fields) from both simulation methods need to be adequately communicated from/to $\Omega_\text{high}$ and $\Omega_\text{low}$, and they must be subsequently updated in the respective regions. This is similar to simulation methods of multiphysics problems~\cite{Bungartz2006,Perelman1961}. Eventually, once these values align, we obtain a converged complete solution \textphi{}. In the proposed method, this is accomplished using an iterative method, i.e., a method that tries to find a local solution (fixed point) iteratively~\cite{Hadjidimos2000,Kelley2003}. More precisely, 

\begin{equation}
    q^n = f(q^{n-1}),
\end{equation}

\noindent where, in this case, $q^n$ is the quantity on \(\Gamma\) that is communicated between $\Omega_\text{low}$ and $\Omega_\text{high}$ at timestep $n$, and $f(q)$ is the update function, given by the iterative method. The stability and convergence characteristics of the iterative method highly depend on the quality of the initial approximate solution, i.e., the {initial condition} $q^0$. 

To find the initial condition $q^0$, the complete network is first solved completely using the MNA, i.e., the high abstraction method. In \cref{fig:ExampleNetwork}, $\Omega_\text{low}$ consists of crossings and \mbox{T-junctions} but could, in reality, contain \emph{{any}} arbitrary shape that can be simulated using CFD. Therefore, the regions in $\Omega_\text{low}$ are replaced by fully connected graphs, where each \mbox{in-/outlet} is treated as a node. This replacement is depicted {in} 
 \cref{fig:Figure2}b. The resulting network can be solved using only the MNA. This is \emph{not} the solution \textphi{} to the actual problem, since we abstracted $\Omega_\text{low}$ by fully connected graphs. However, the solution to this abstracted problem can be used as an initial condition $q^0$ for the iterative method.

\begin{figure}[H]
	\begin{subfigure}[b]{0.45\columnwidth}
		\centering
		\includegraphics[width=0.8\textwidth]{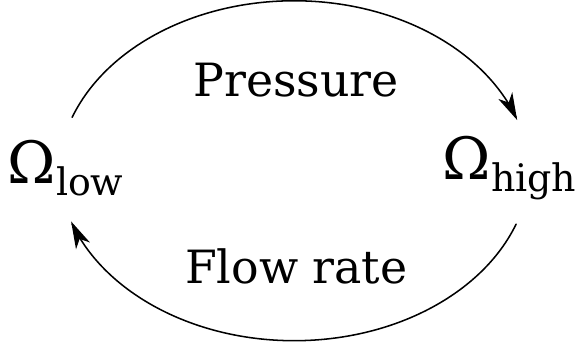}
\captionsetup{justification=centering}
		\caption{ }
		\label{subfig:Source}
	\end{subfigure}
	\hfill
	\begin{subfigure}[b]{0.45\columnwidth}
		\centering
		\includegraphics[width=0.8\textwidth]{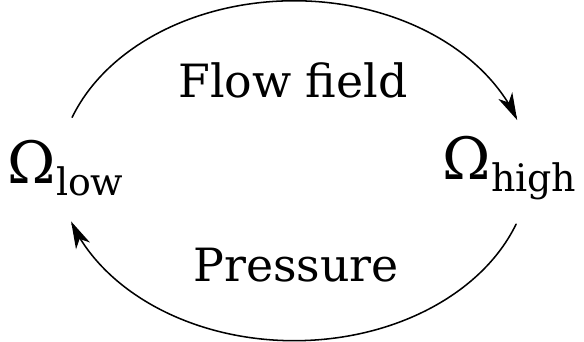}
\captionsetup{justification=centering}
		\caption{ }
		\label{subfig:Sink}
	\end{subfigure}
	\caption{{Communication} 
 schemes for the pressure and flow fields. {(\textbf{a})} 
 Communicate the flow rate from $\Omega_\text{high}$ to $\Omega_\text{low}$ and the pressure vice versa. Here, the flow field information must be extrapolated from the communicated flow rate. (\textbf{b}) Communicate the flow field from $\Omega_\text{low}$ to $\Omega_\text{high}$ and the pressure vice versa.}
\label{fig:Communication}
\end{figure}

Having that, the communication can be performed in either of the two ways sketched in \cref{fig:Communication}. A low abstraction solver (such as the LBM solver) calculates the pressure and velocity fields directly, whereas a high abstraction solver (such as the MNA solver) calculates the flow rate in a channel rather than the velocity field. This means that if we map the flow rate from the MNA solver to the LBM solver (Figure \ref{fig:Communication}a), we need to extrapolate the flow profile based on the flow rate, whereas the reverse mapping (\cref{fig:Communication}b) can be done directly (provided that \textGamma{} is located sufficiently far in a straight channel section, such that the pressure is sufficiently uniform on the channel cross-section and can be treated as a point value). Regions that belong to~$\Omega_\text{high}$ but are not connected to ground nodes (the green {nodes in} 
 \cref{fig:Figure2}a) need to communicate according to \cref{fig:Communication}a in at least one node. If this is not the case, the absolute pressure, i.e., the pressure difference with respect to the reference pressure $p^0$ at the ground nodes, is not propagated correctly. These regions are highlighted in {red in} 
 \cref{fig:Figure2}a).

To ensure a converged complete solution, we used an iterative method based on \emph{Successive Over-Relaxation} (SOR,~\cite{Hadjidimos2000}) 
 to find the values for pressure and velocity on $\Gamma$. From the initial condition $q^0$, the boundary conditions of the LBM solver are updated in every iteration according to the newly found values from the MNA, corrected by a {relaxation factor}~\textalpha{}, i.e., 

 \begin{equation}\label{eq:Iterative}
     q^{n}_\text{low} = (1-\alpha) \; q^{n-1}_\text{low} + \alpha \; q^{n-1}_\text{high}.
 \end{equation}

\noindent
{Here,} 
 $q^n_\text{high}$ is evaluated using the MNA with the most recent pressure and velocity information obtained from the LBM solver. The LBM is in itself also an iterative solver, and the frequency at which the boundary conditions are updated influences the stability of the LBM. To ensure stability in the LBM, the values of the boundary conditions are only updated every $\theta$ \mbox{\enquote{LBM iteration steps}}. The iterative approach in \cref{eq:Iterative} is solved until the convergence criterion

\begin{equation}
     |q^n - q^{n-1}| \leq \epsilon
\end{equation}

\noindent is met, where $\epsilon$ can be chosen arbitrarily small (until machine precision is reached).

\section{Case Studies and Evaluation Results}
\label{sec:Results}

The approach to accelerate CFD simulations, as proposed above, has been implemented for the considered continuous channel-based microfluidics. As corresponding simulation tools, we used OpenLB v1.5 \cite{OLB2021} for the LBM and an \mbox{in-house} implementation of the MNA (which, as discussed above, are used as the low and high abstraction simulation methods, respectively). The source code of the proposed method, that has been developed for this work, is available online (available at: \url{https://github.com/cda-tum/mmft-hybrid-simulator}{; accessed on 6 October 2023}) 
. 

Using the resulting implementation, several case studies were conducted to evaluate whether the idea proposed in this work indeed yields an improvement in CFD simulations. This section summarizes the respectively obtained findings. To this end, first the setup of the case studies is reviewed. Afterward, the obtained results are presented and discussed.

\subsection{Setup: Considered Cases and Parameters}
\label{sec:Example}

\begin{figure}[t]
    \hspace{0.12\columnwidth}
     \begin{subfigure}[b]{0.24\columnwidth}
         \centering
         \includegraphics[width=1.0\textwidth]{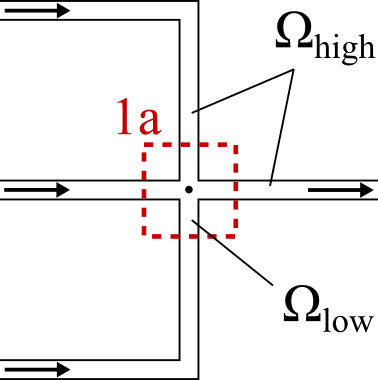}
         \caption{Network 1}
         \label{fig:ExampleNetwork1}
     \end{subfigure}
     \hspace{0.12\columnwidth}
     \hspace{0.04\columnwidth}
     \begin{subfigure}[b]{0.36\columnwidth}
         \centering
         \includegraphics[width=1.0\textwidth]{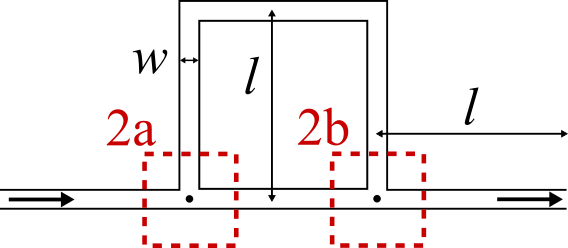}
         \caption{Network 2}
         \label{fig:ExampleNetwork2}
     \end{subfigure}
     \\
     \hfill\\
     \begin{subfigure}[b]{0.48\columnwidth}
         \centering
         \includegraphics[width=1.0\textwidth]{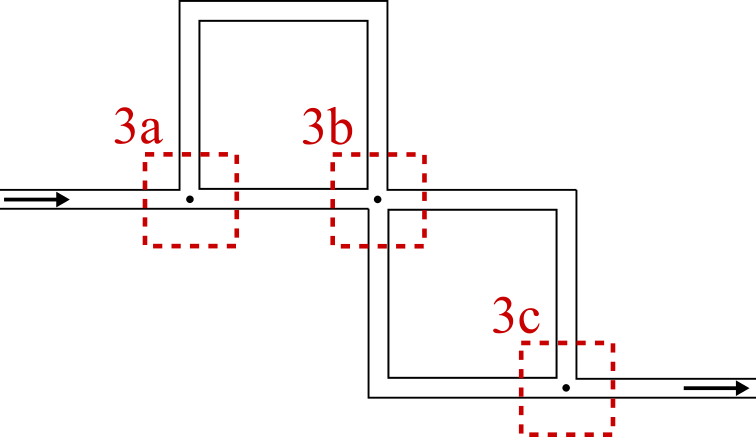}
         \caption{Network 3}
         \label{fig:ExampleNetwork3}
     \end{subfigure}
     \hfill
     \begin{subfigure}[b]{0.48\columnwidth}
         \centering
         \includegraphics[width=1.0\textwidth]{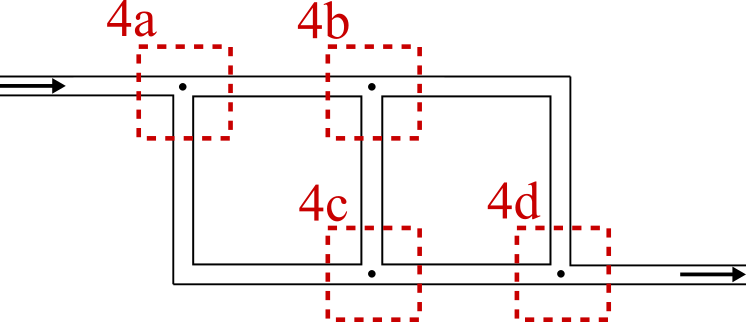}
         \caption{Network 4}
         \label{fig:ExampleNetwork4}
     \end{subfigure}
        \caption{The networks of the considered case studies. The number of separate regions in $\Omega_\text{low}$ increments with the case studies, starting at one region (\textbf{a}) for Network 1 and ending at four regions (\textbf{a}--\textbf{d}) for Network 4.}
        \label{fig:PaperNetworks}
\end{figure}

In our case studies, four different continuous \mbox{channel-based} microfluidic networks, as shown in Figure \ref{fig:PaperNetworks} (denoted {Network 1--4} 
 in the following), were considered. All networks are \mbox{two-dimensional}. For each of these networks, we considered different amounts of disconnected regions in \mbox{$\Omega_\text{low}$}, as well as different lengths of the channels that constitute \mbox{$\Omega_\text{high}$}---providing a proper variety of test cases with different coverages of $\Omega_\text{low}$- and $\Omega_\text{high}$-regions. 
More precisely, following the discussion from \cref{sec:Application}, all junctions and crossings of channels are identified as $\Omega_{low}$-region (indicated by the red dotted squares in \cref{fig:PaperNetworks}), and the connecting channels are identified as $\Omega_{high}$. \textGamma{} was always placed inside a straight channel at a distance of two channel thicknesses from the corresponding junction or crossing. The length~$l$ of all channels is subsequently set to 1, 2, 3, and 4 mm for all four networks, whereas the $\Omega_\text{low}$-regions remain constant. Overall, this leads to a total of 16 separate cases.

The channels of all networks are rectangular with a width of 100~\textmu m, and all networks have the inlets located on the \mbox{left-hand} side, with pressure boundary conditions of 1000 Pa, and outlets on the \mbox{right-hand} side, with pressure boundary conditions of 0 Pa. For all test cases, the fluid inside the network is an incompressible homogeneous fluid with a density of 1000 kg/m\(^\text{3}\) and a kinematic viscosity of 1\(\cdot\)10$^{-6}$  m\(^\text{2}\)/s.

The test cases are solved on a regular grid with a resolution of 20 grid cells across the width of each channel, where applicable (i.e., the resolution of the CFD simulation and $\Omega_\text{low}$). Additionally, for the proposed method, we set the values $\theta = 10$, $\epsilon = 0.01$, and $\alpha = 0.01$ for Networks 1 and 2, and $0.003$ for Networks 3 and 4. 
All simulations were performed without compiler optimization on a single CPU core (no parallelism) of an AMD Ryzen Threadripper PRO 5955WX CPU \cite{ThreadripperPro5955WX}.

\subsection{Obtained Results}

In order to evaluate the performance of the proposed method, two aspects are essential: The runtime required for the respective CFD simulations (as we are aiming to accelerate them), as well as the accuracy (as potential accelerations ideally should yield the same results). 

Concerning the former, Table \ref{tab:Runtimes} lists the respectively obtained results. Here, for all networks from \cref{fig:PaperNetworks} (listed from left to right), as well as for all channel lengths (listed in the rows), the correspondingly required runtimes of the original as well as the proposed method are provided. Additionally, the resulting speed-ups obtained by the proposed method are listed. As mentioned previously, the LBM was used in this work as a representative for CFD simulation methods. However, the computational complexity of the LBM is similar to that of, e.g., the FVM \cite{Takken2022} and speed-ups of similar order of magnitude can be expected for other simulation methods.

\begin{figure*}[h!]
          \begin{subfigure}[b]{0.48\textwidth}
         \centering
         \includegraphics[width=1.0\textwidth]{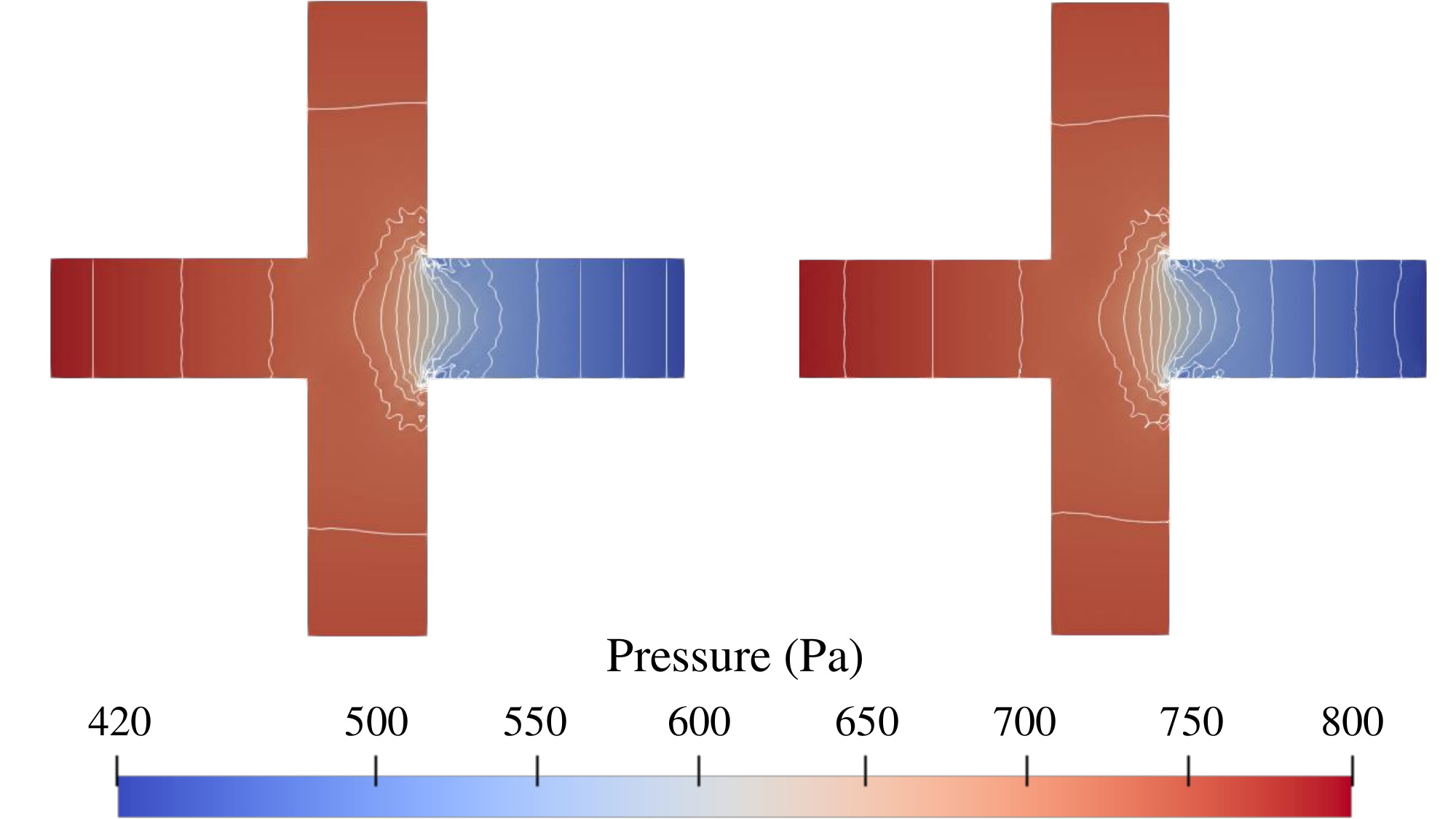}
         \caption{Pressure field of Region \emph{1a} with \(l\) is 1.}
         \label{fig:1a_cross0_pressure}
     \end{subfigure}
     \hfill
     \begin{subfigure}[b]{0.48\textwidth}
         \centering
         \includegraphics[width=1.0\textwidth]{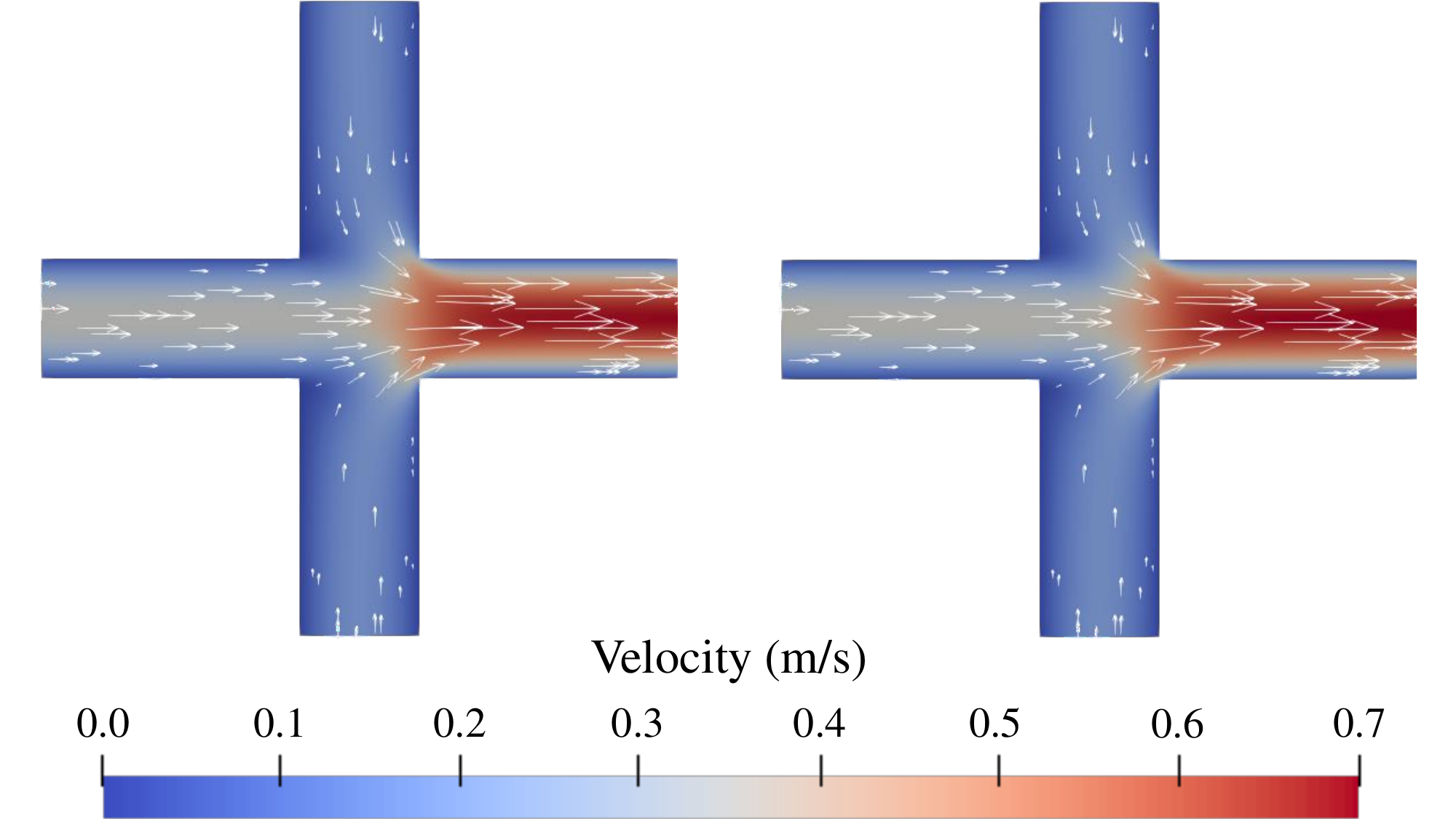}
         \caption{Velocity field of Region \emph{1a}  with \(l\) is 1.}
         \label{fig:1a_cross0_velocity}
     \end{subfigure}\\
    \begin{subfigure}[b]{0.48\textwidth}
         \centering
         \includegraphics[width=1.0\textwidth]{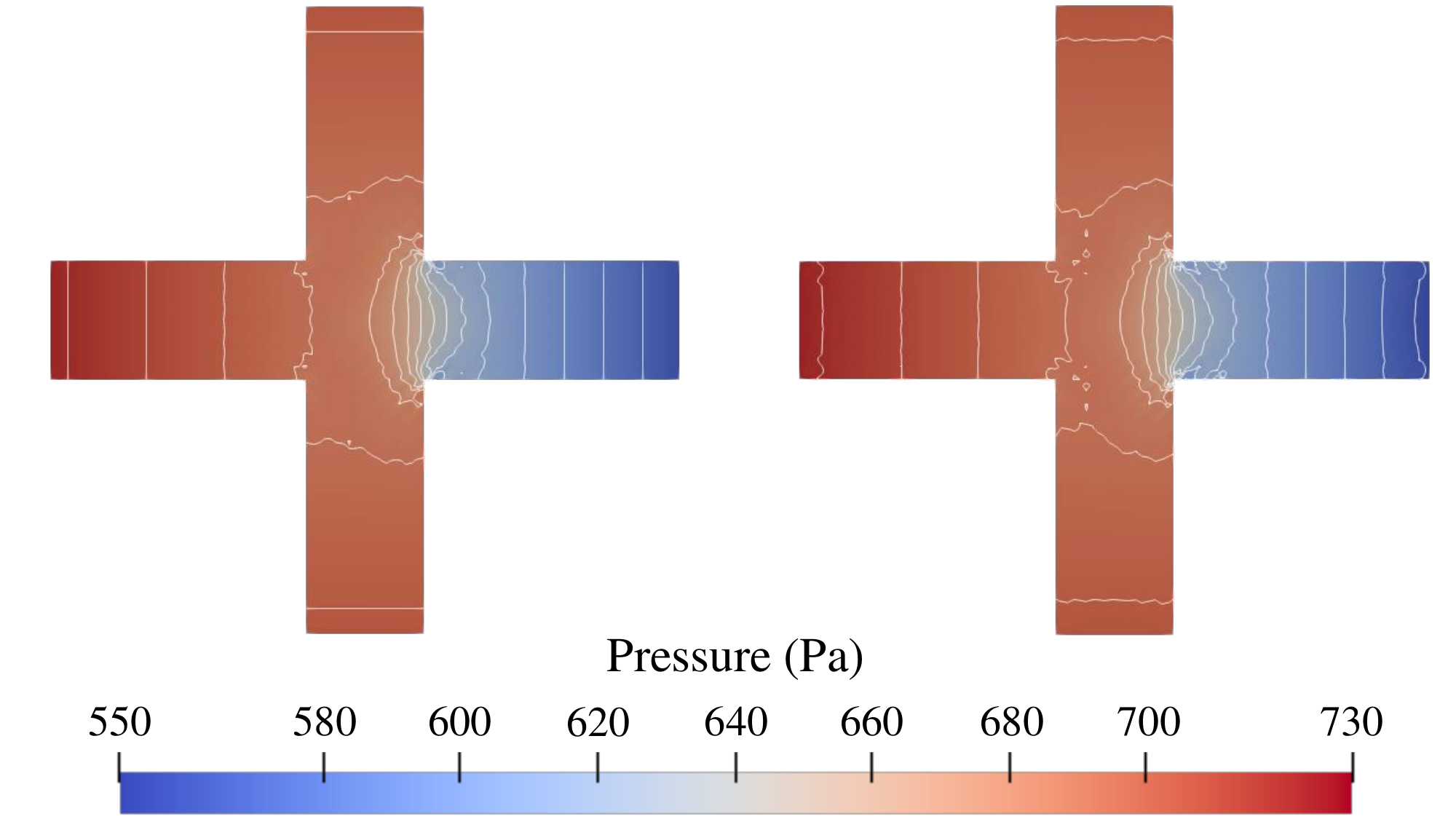}
         \caption{Pressure field of Region \emph{1a} with \(l\) is 2.}
         \label{fig:1b_cross0_pressure}
        \end{subfigure}
     \hfill
     \begin{subfigure}[b]{0.48\textwidth}
         \includegraphics[width=1.0\textwidth]{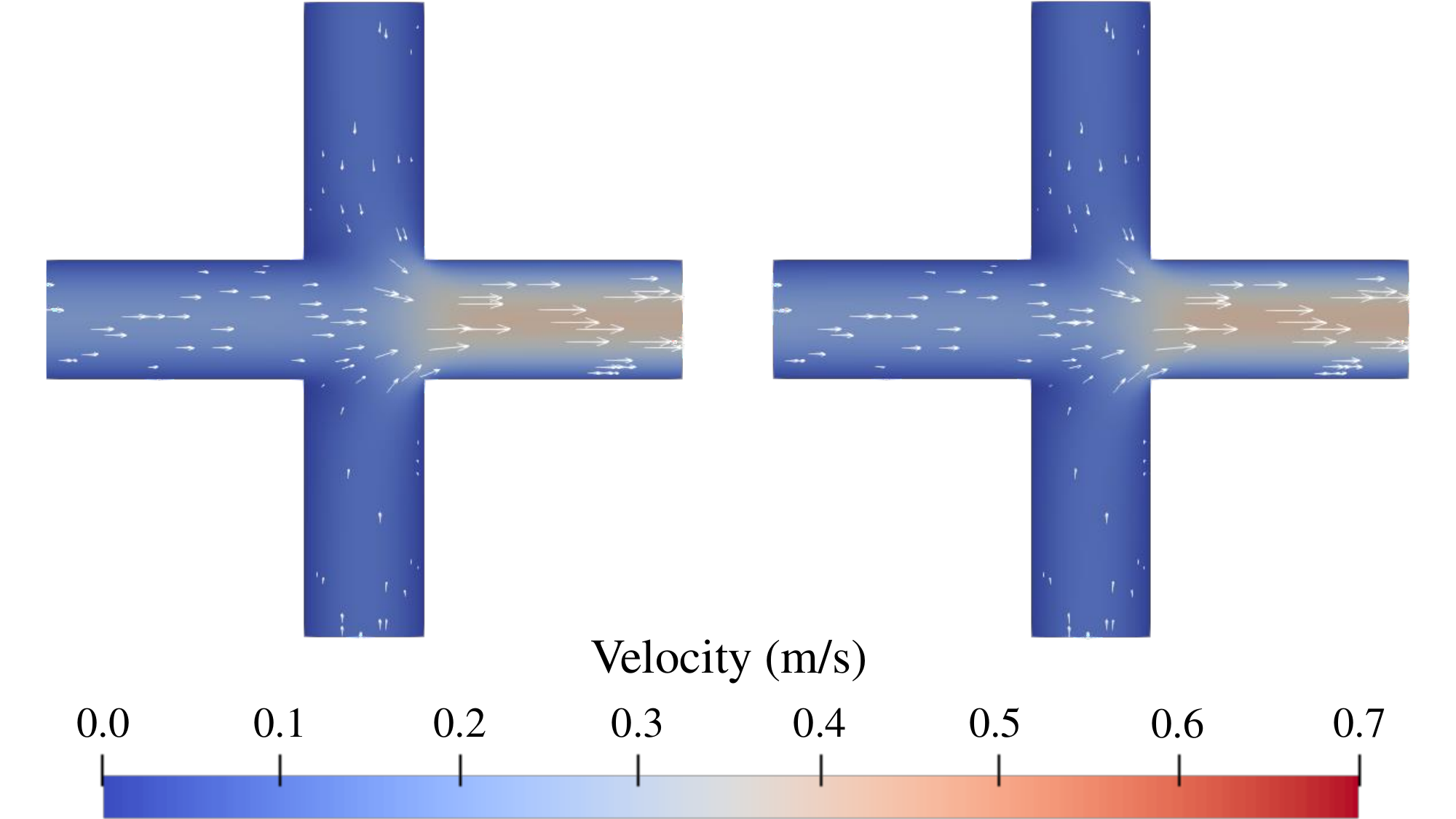}
         \caption{Velocity field of Region \emph{1a} with \(l\) is 2.}
         \label{fig:1b_cross0_velocity}
     \end{subfigure}\\
     \\
    \begin{subfigure}[b]{0.48\textwidth}
         \centering
         \includegraphics[width=1.0\textwidth]{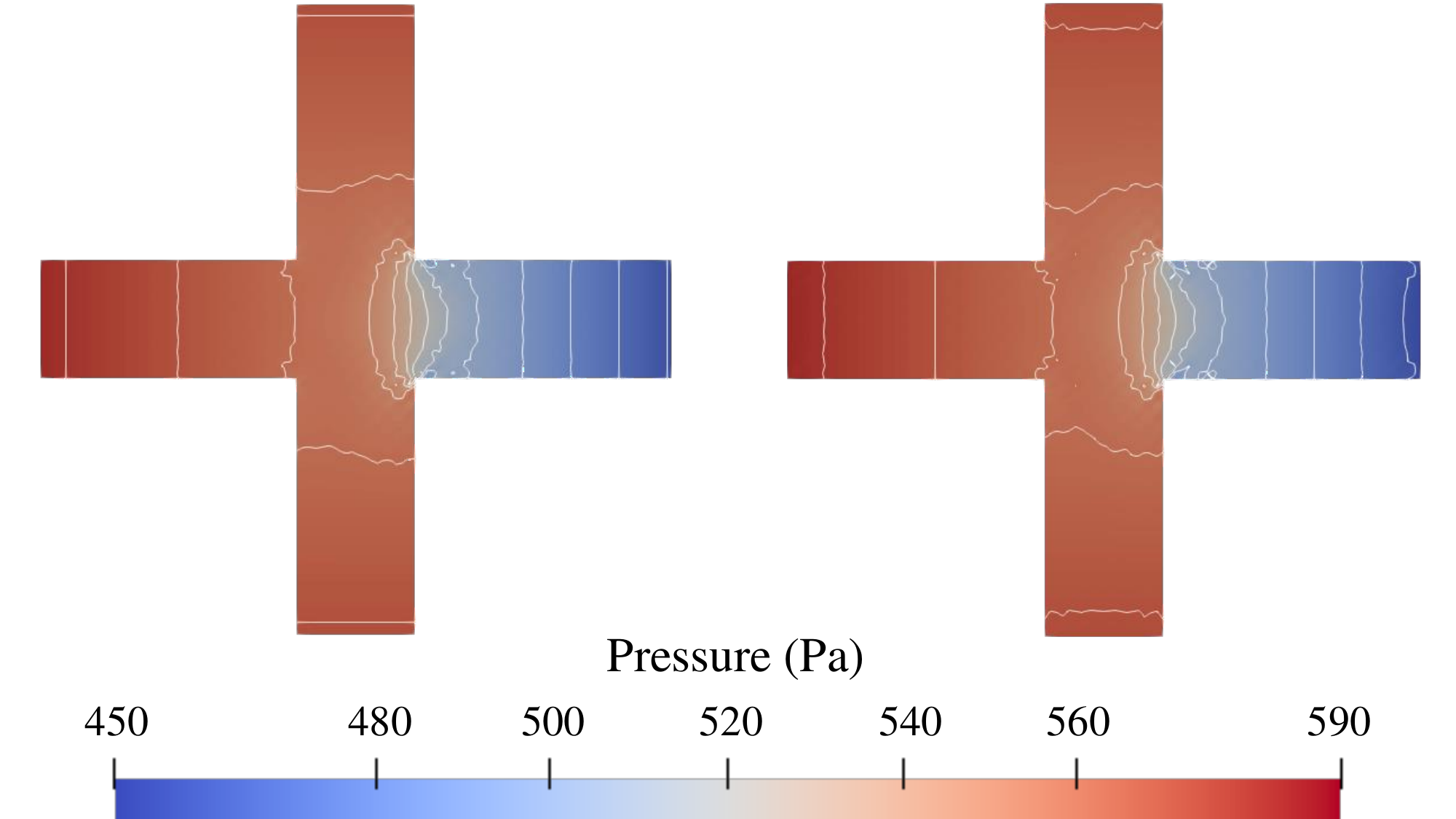}
         \caption{Pressure field of Region \emph{1a} with \(l\) is 3.}
         \label{fig:1c_cross0_pressure}
        \end{subfigure}
     \hfill
     \begin{subfigure}[b]{0.48\textwidth}
         \includegraphics[width=1.0\textwidth]{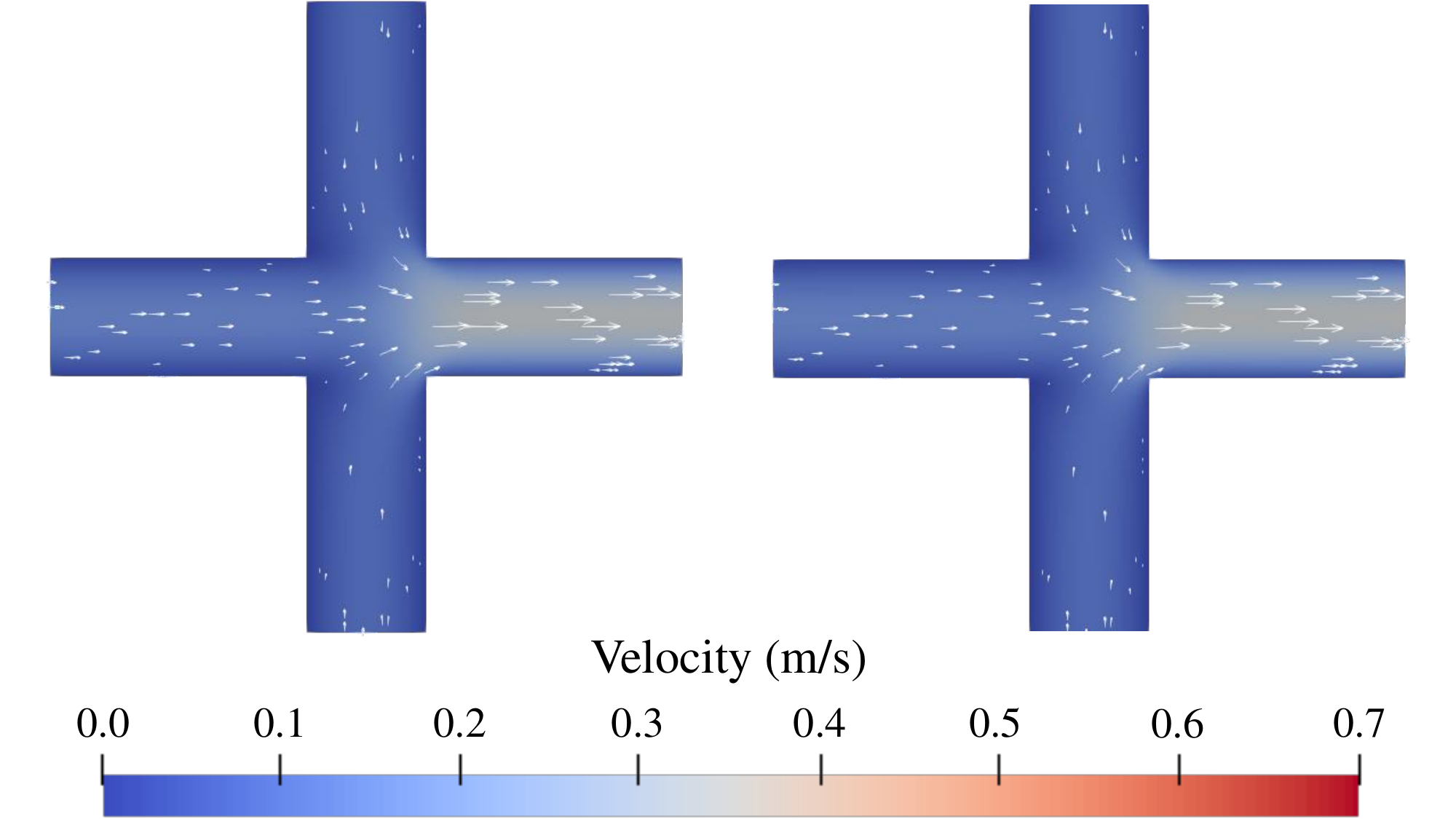}
         \caption{Velocity field of Region \emph{1a} with \(l\) is 3.}
         \label{fig:1c_cross0_velocity}
     \end{subfigure}\\
     \\
    \begin{subfigure}[b]{0.48\textwidth}
         \centering
         \includegraphics[width=1.0\textwidth]{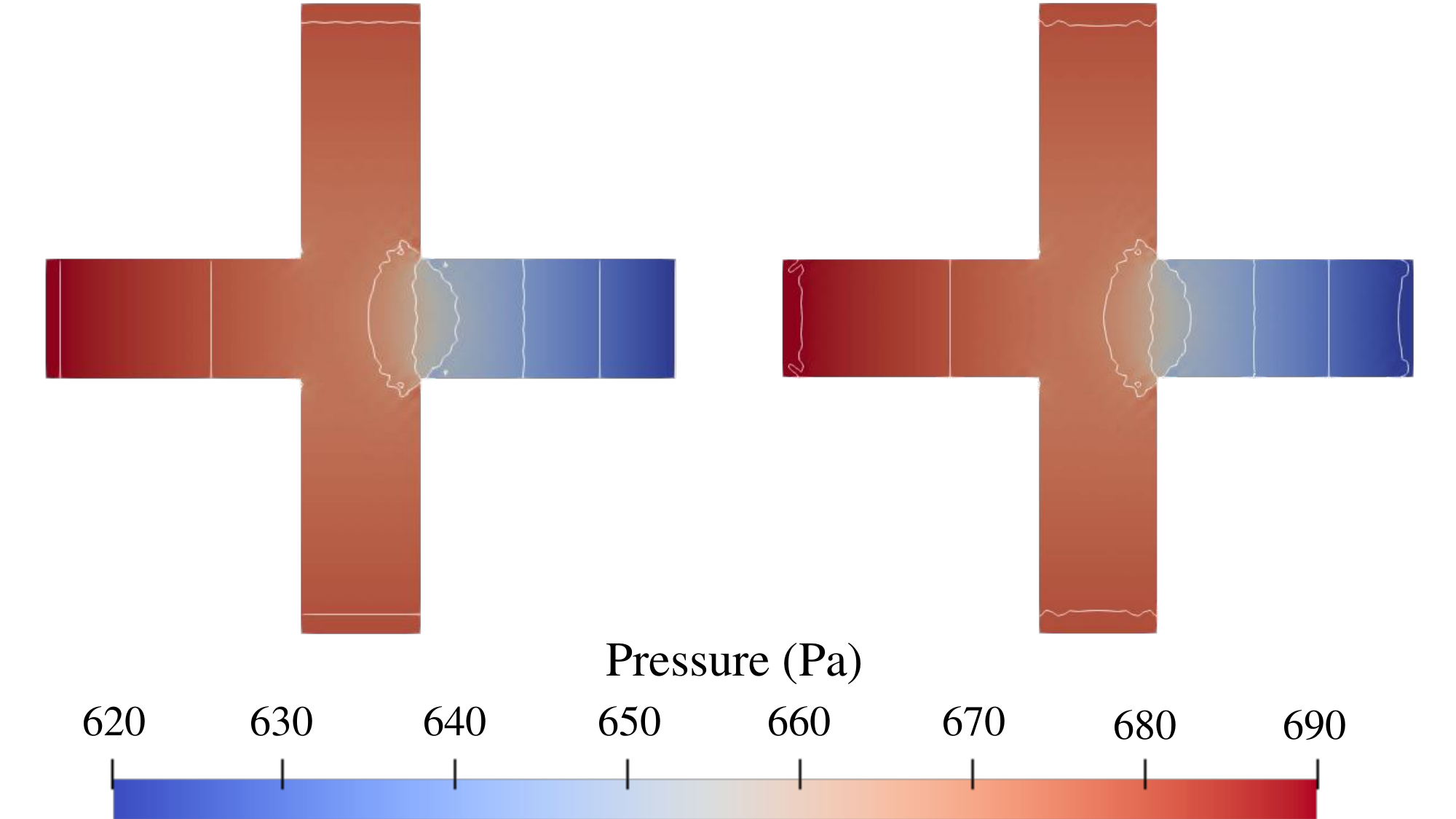}
         \caption{Pressure field of Region \emph{1a} with \(l\) is 4.}
         \label{fig:1d_cross0_pressure}
        \end{subfigure}
     \hfill
     \begin{subfigure}[b]{0.48\textwidth}
         \includegraphics[width=1.0\textwidth]{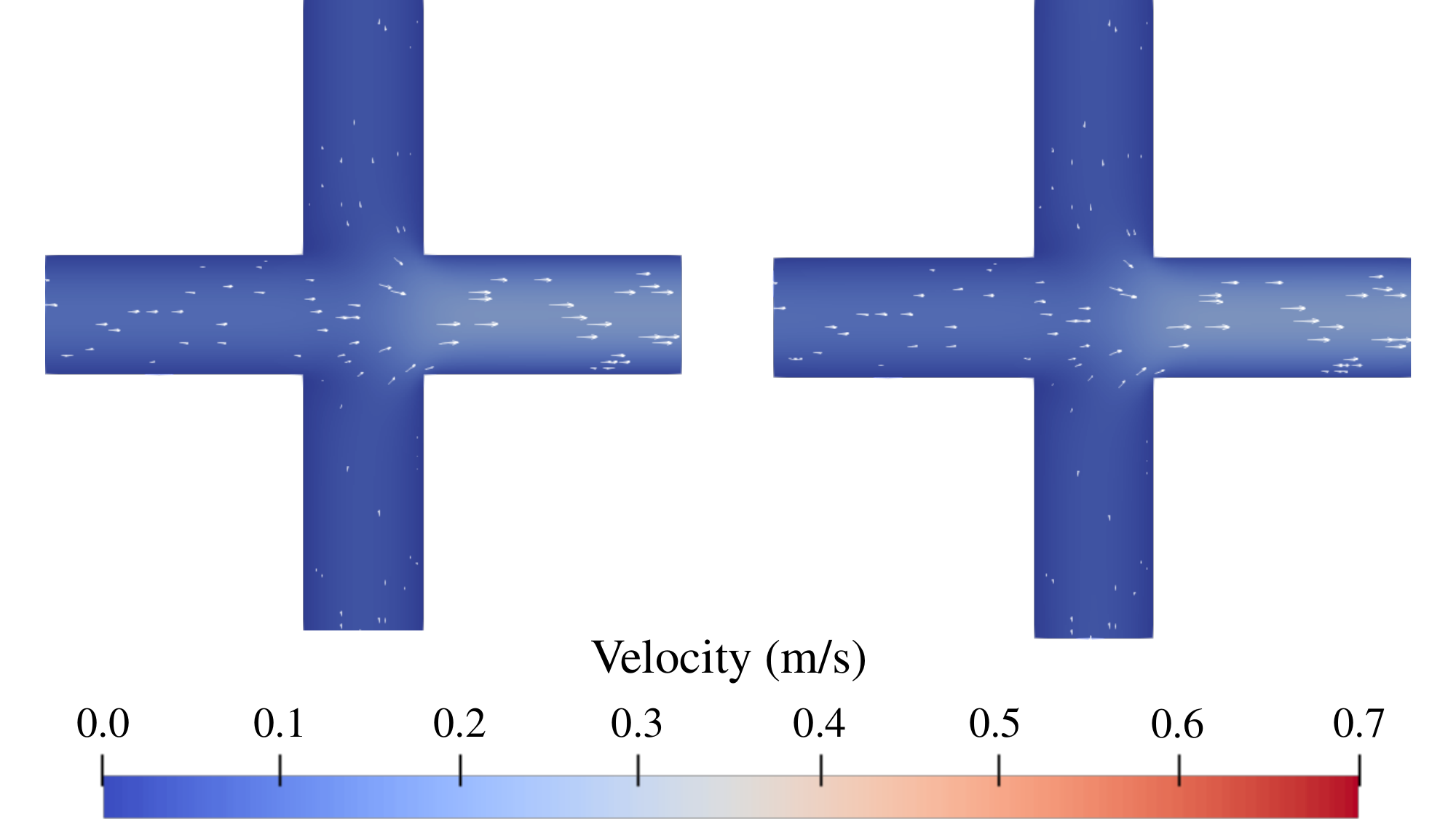}
         \caption{Velocity field of Region \emph{1a} with \(l\) is 4.}
         \label{fig:1d_cross0_velocity}
     \end{subfigure}\\
        \caption{{The} 
 {pressure} 
 and velocity fields obtained from the CFD simulations (\textbf{left}) and the proposed method (\textbf{right}) for Region \emph{1a} in Network 1 with \(l\text{ is }1,2,3\text{ and }4\). }
        \label{fig:1_cross0}
\end{figure*}

\begin{figure*}[h!]
     \begin{subfigure}[b]{0.48\textwidth}
         \centering
         \includegraphics[width=1.0\textwidth]{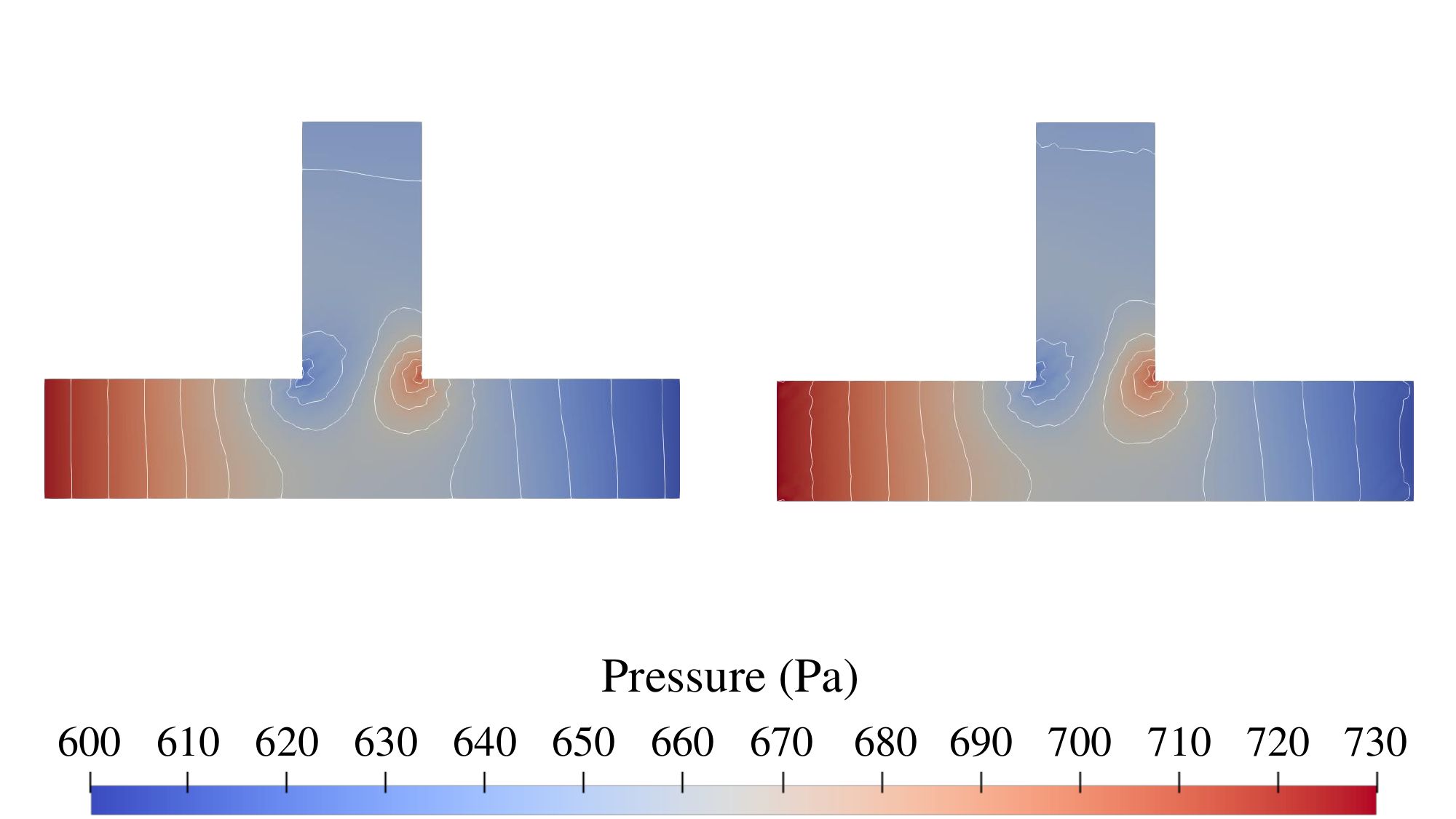}
         \caption{Pressure field of Region \emph{2a}.}
         \label{fig:2a_T0_pressure}
     \end{subfigure}
     \hfill
     \begin{subfigure}[b]{0.48\textwidth}
         \centering
         \includegraphics[width=1.0\textwidth]{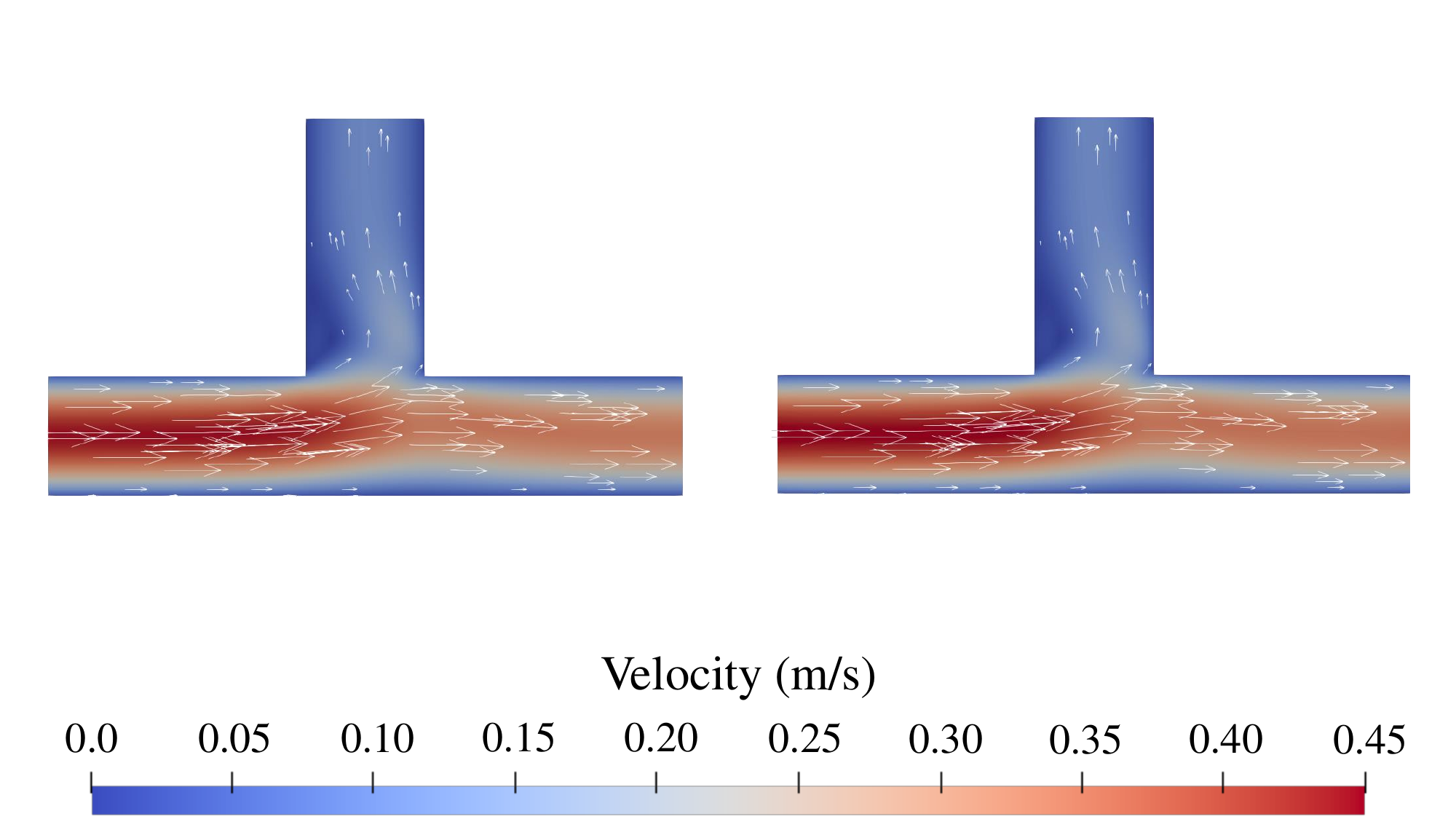}
         \caption{Velocity field of Region \emph{2a}.}
         \label{fig:2a_T0_velocity}
     \end{subfigure}\\
     \\
    \begin{subfigure}[b]{0.48\textwidth}
         \centering
         \includegraphics[width=1.0\textwidth]{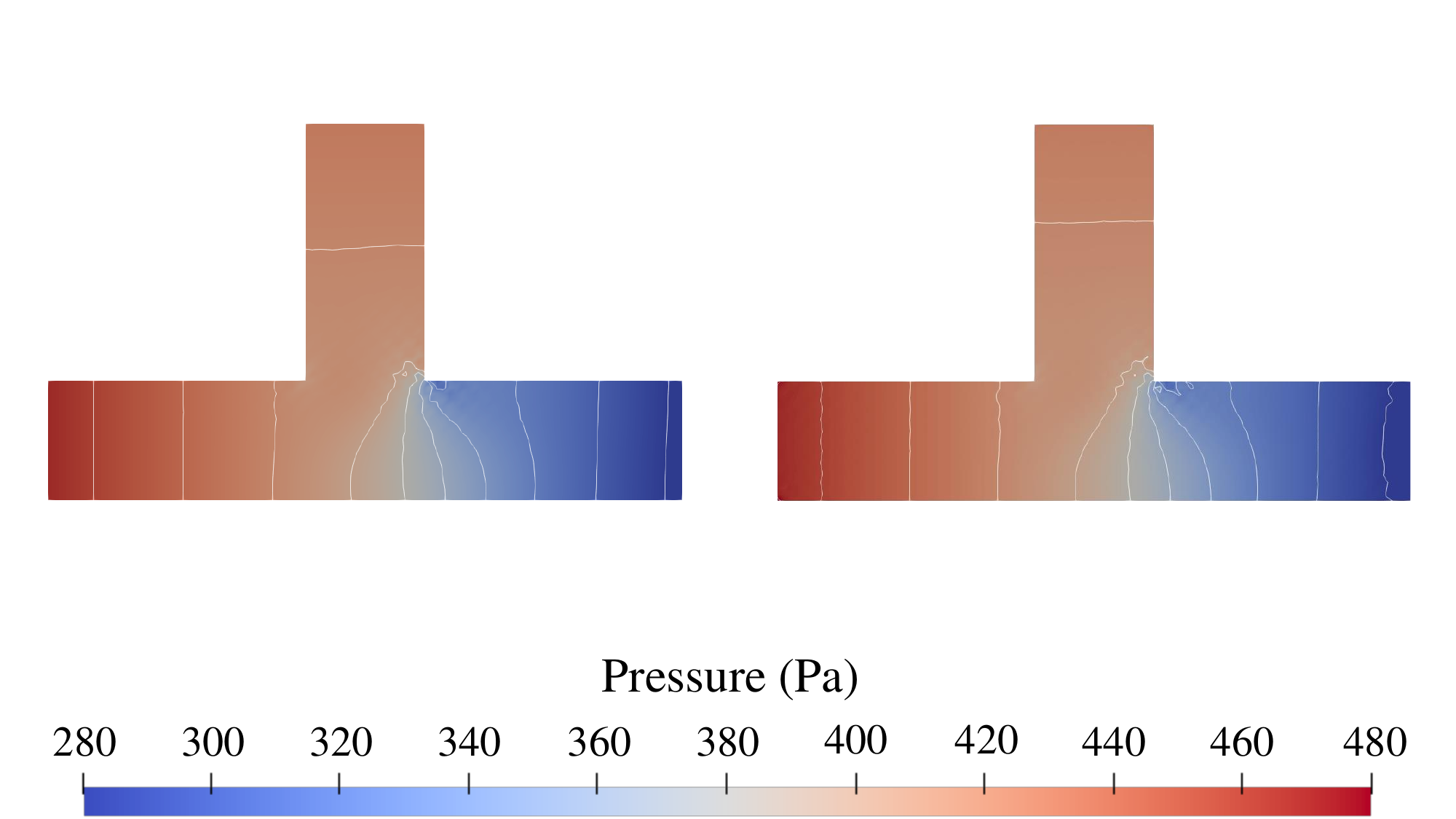}
         \caption{Pressure field of Region \emph{2b}.}
         \label{fig:2a_T1_pressure}
     \end{subfigure}
     \hfill
     \begin{subfigure}[b]{0.48\textwidth}
         \centering
         \includegraphics[width=1.0\textwidth]{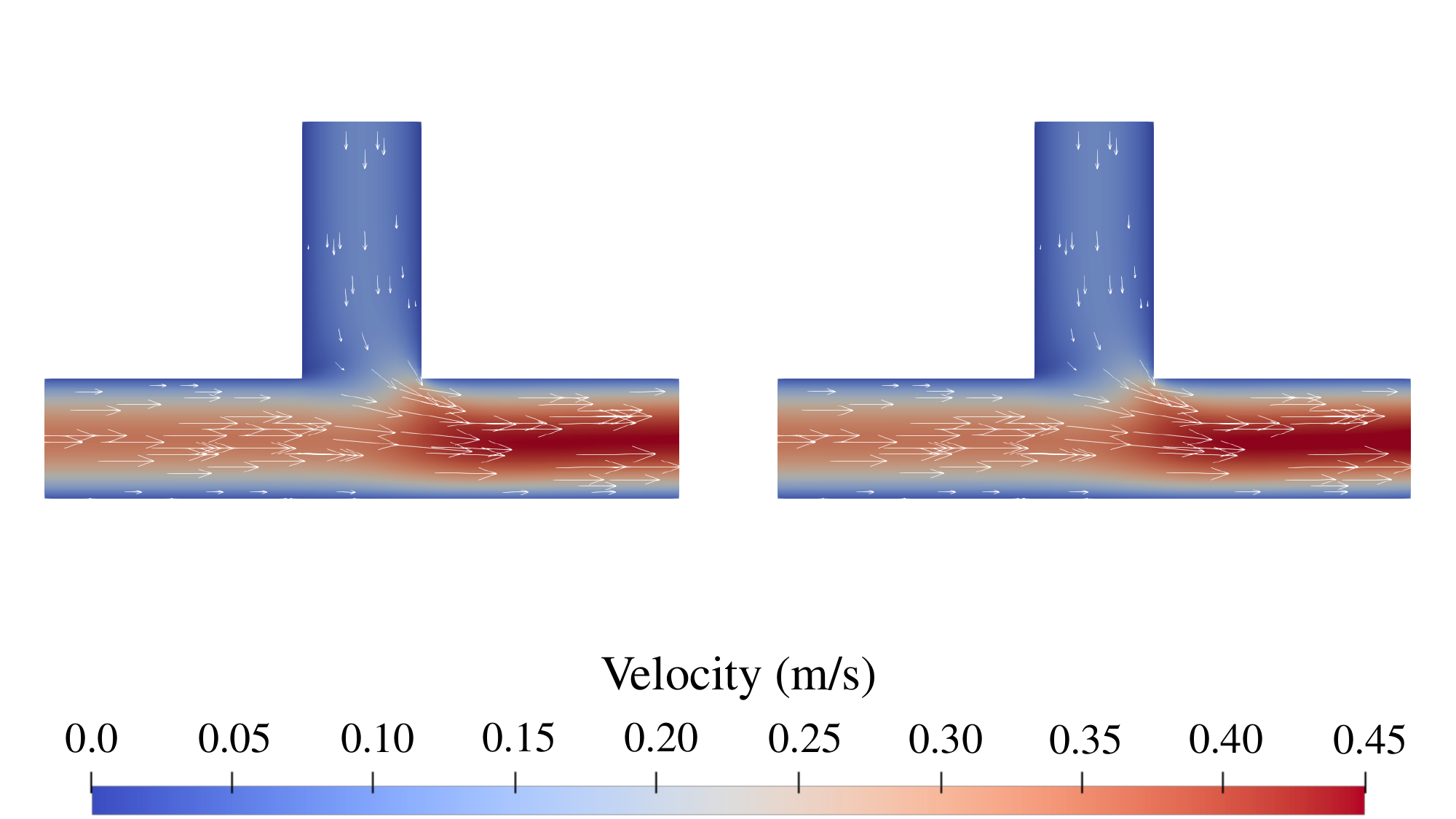}
         \caption{Velocity field of Region \emph{2b}.}
         \label{fig:2a_T1_velocity}
     \end{subfigure}
        \caption{{The} 
 pressure and velocity fields obtained from the CFD simulations (\textbf{left}) and the proposed method (\textbf{right}) for Regions \emph{2a} and \emph{2b} in Network 2 at \(l\text{ is }1\). }
        \label{fig:2a}
\end{figure*}

\begin{figure*}[h!]
     \begin{subfigure}[b]{0.48\textwidth}
         \centering
         \includegraphics[width=1.0\textwidth]{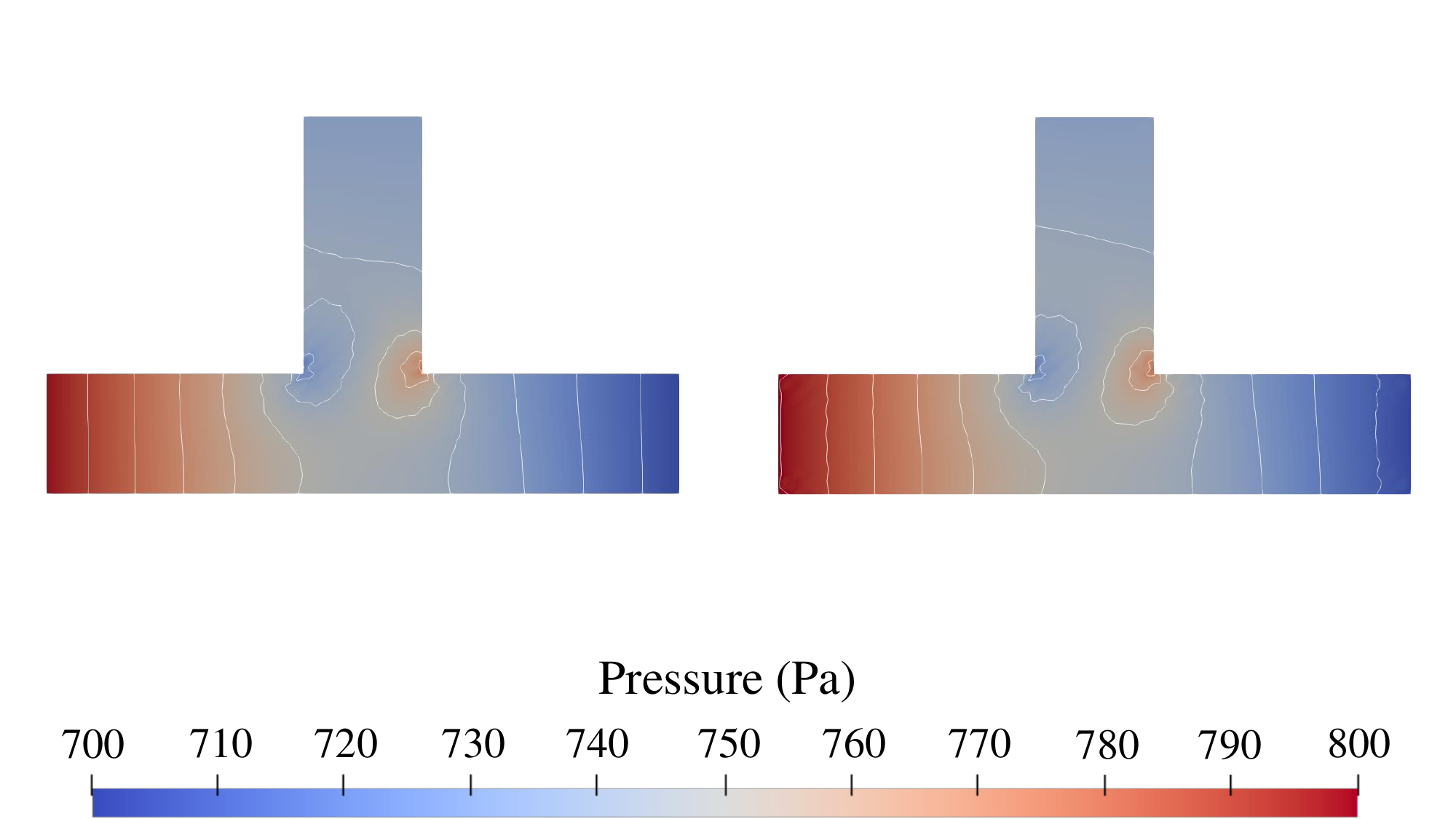}
         \caption{Pressure field of Region \emph{3a}.}
         \label{fig:3a_T0_pressure}
     \end{subfigure}
     \hfill
     \begin{subfigure}[b]{0.48\textwidth}
         \centering
         \includegraphics[width=1.0\textwidth]{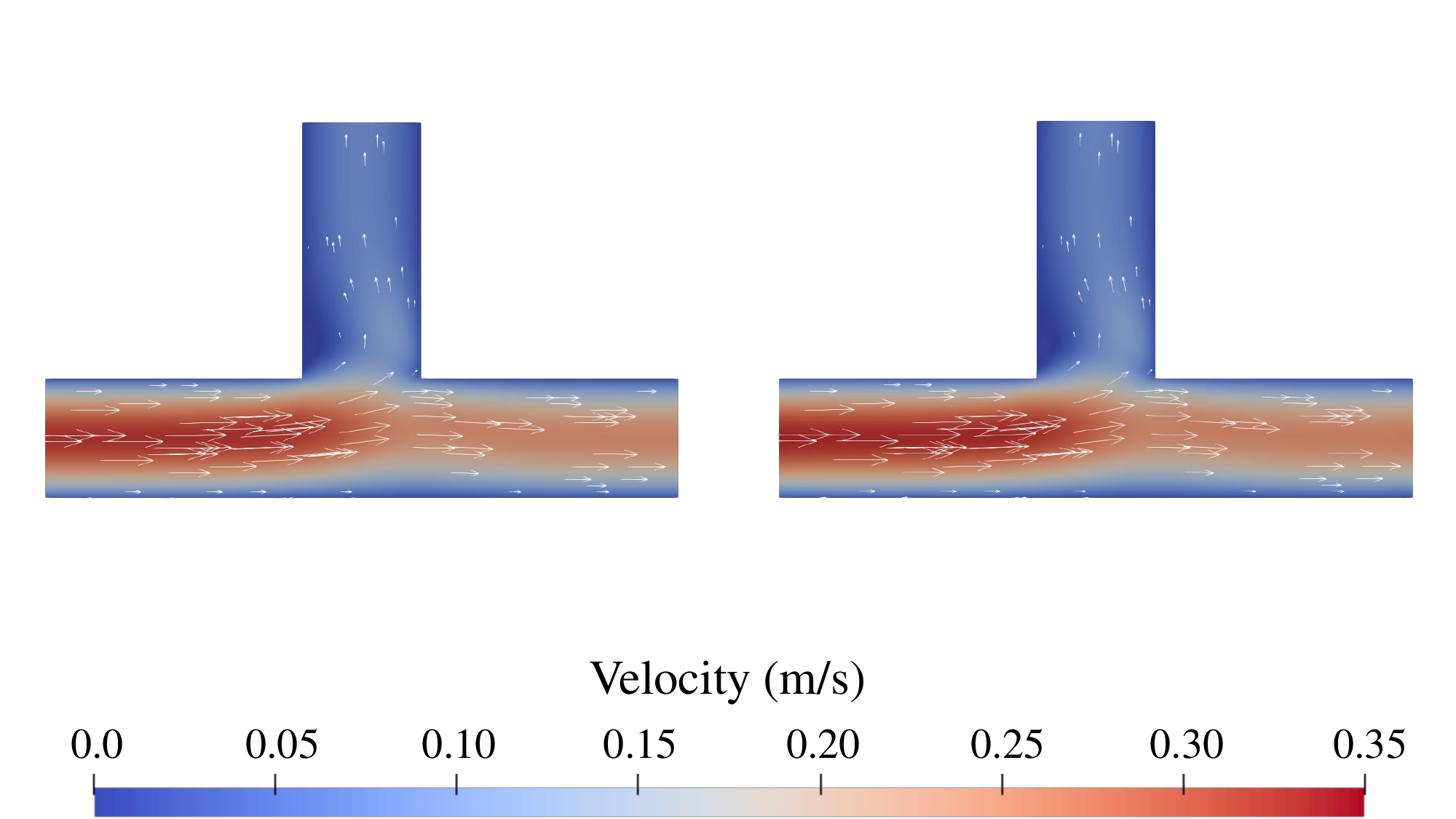}
         \caption{Velocity field of Region \emph{3a}.}
         \label{fig:3a_T0_velocity}
     \end{subfigure}\\
     \\
    \begin{subfigure}[b]{0.48\textwidth}
         \centering
         \includegraphics[width=1.0\textwidth]{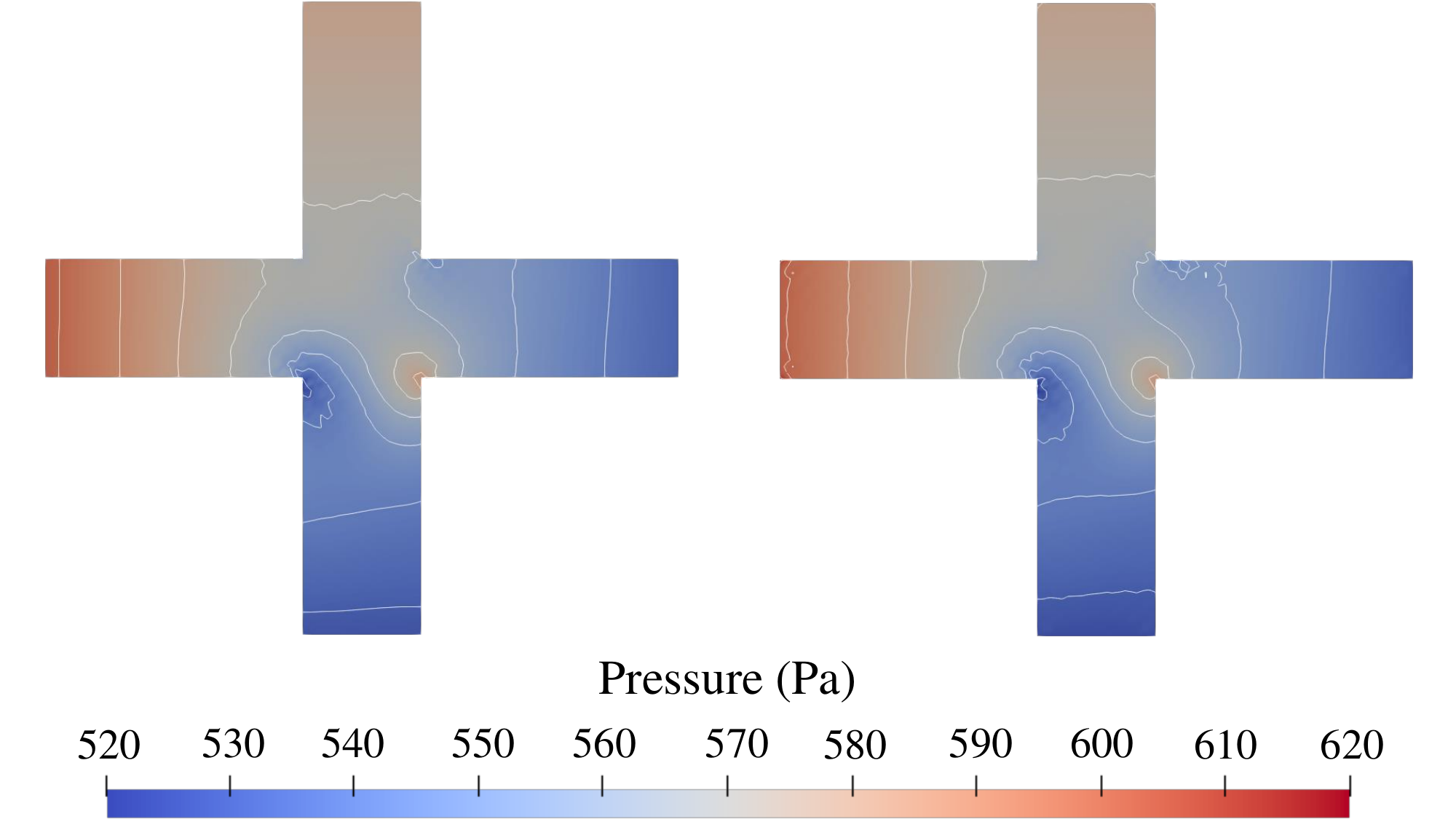}
         \caption{Pressure field of Region \emph{3b}.}
         \label{fig:3a_cross1_pressure}
     \end{subfigure}
     \hfill
     \begin{subfigure}[b]{0.48\textwidth}
         \centering
         \includegraphics[width=1.0\textwidth]{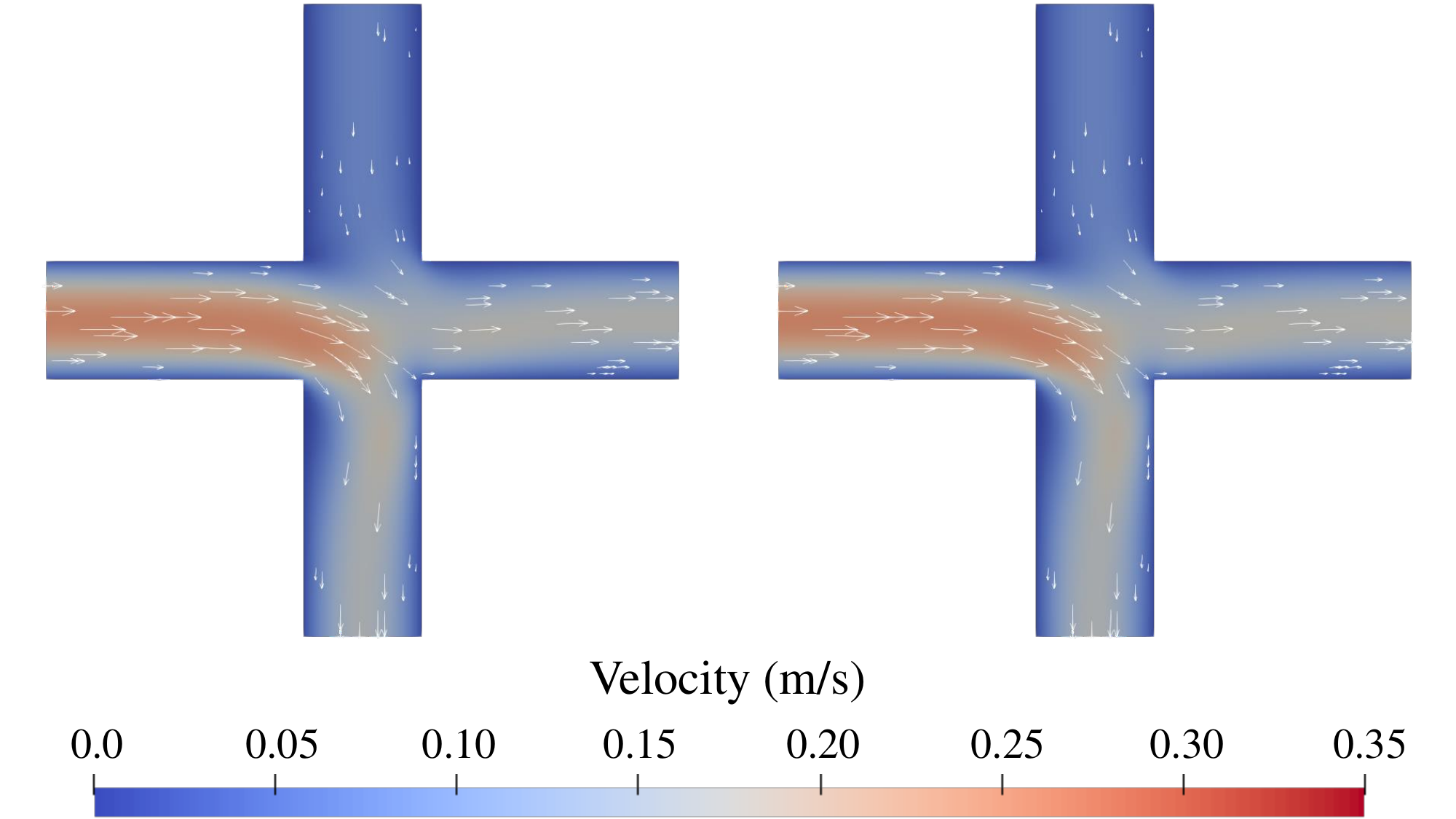}
         \caption{Velocity field of Region \emph{3b}.}
         \label{fig:3a_cross1_velocity}
     \end{subfigure}\\
     \\
    \begin{subfigure}[b]{0.48\textwidth}
         \centering
         \includegraphics[width=1.0\textwidth]{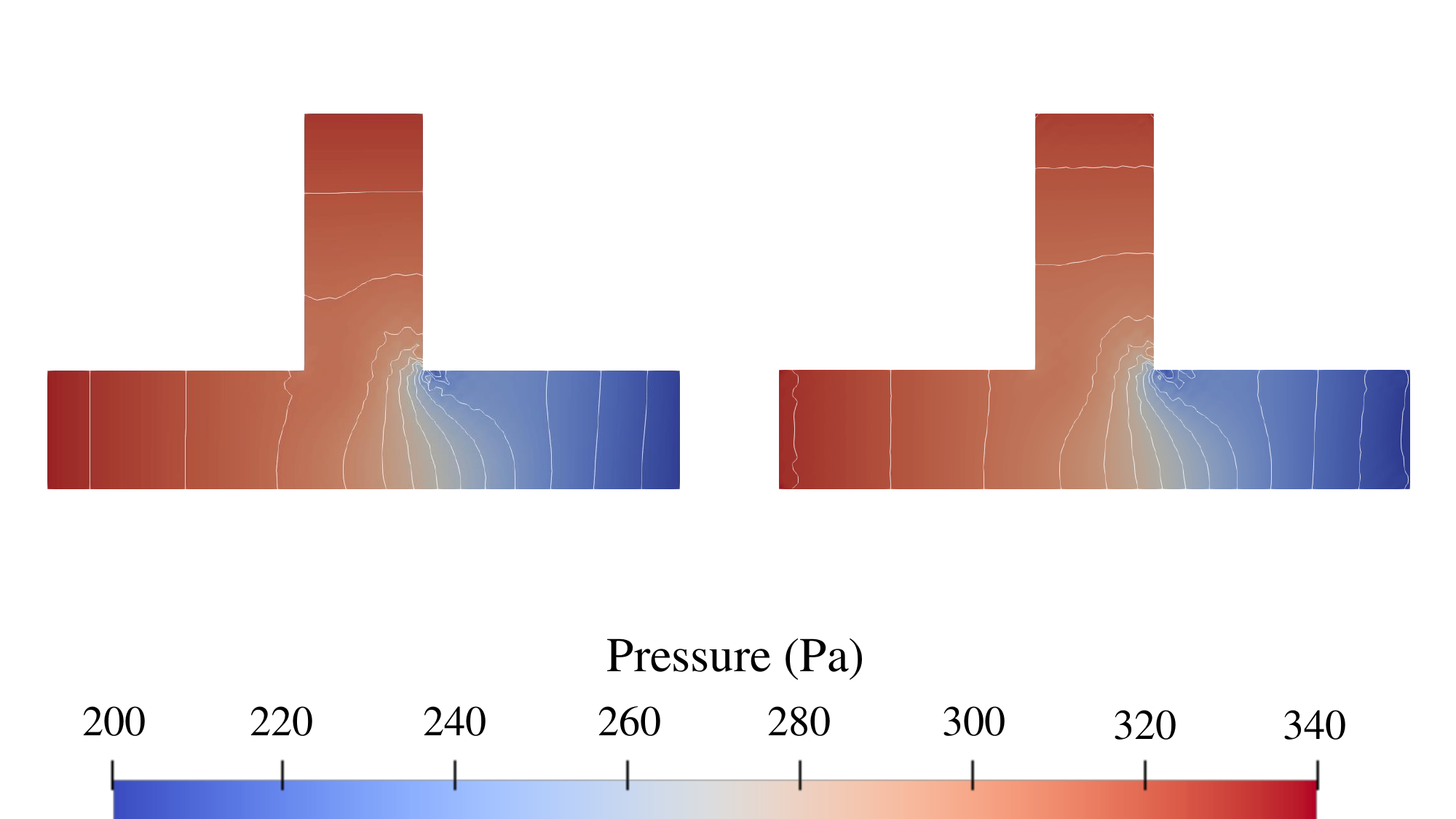}
         \caption{Pressure field of Region \emph{3c}.}
         \label{fig:3a_T2_pressure.pdf}
     \end{subfigure}
     \hfill
     \begin{subfigure}[b]{0.48\textwidth}
         \centering
         \includegraphics[width=1.0\textwidth]{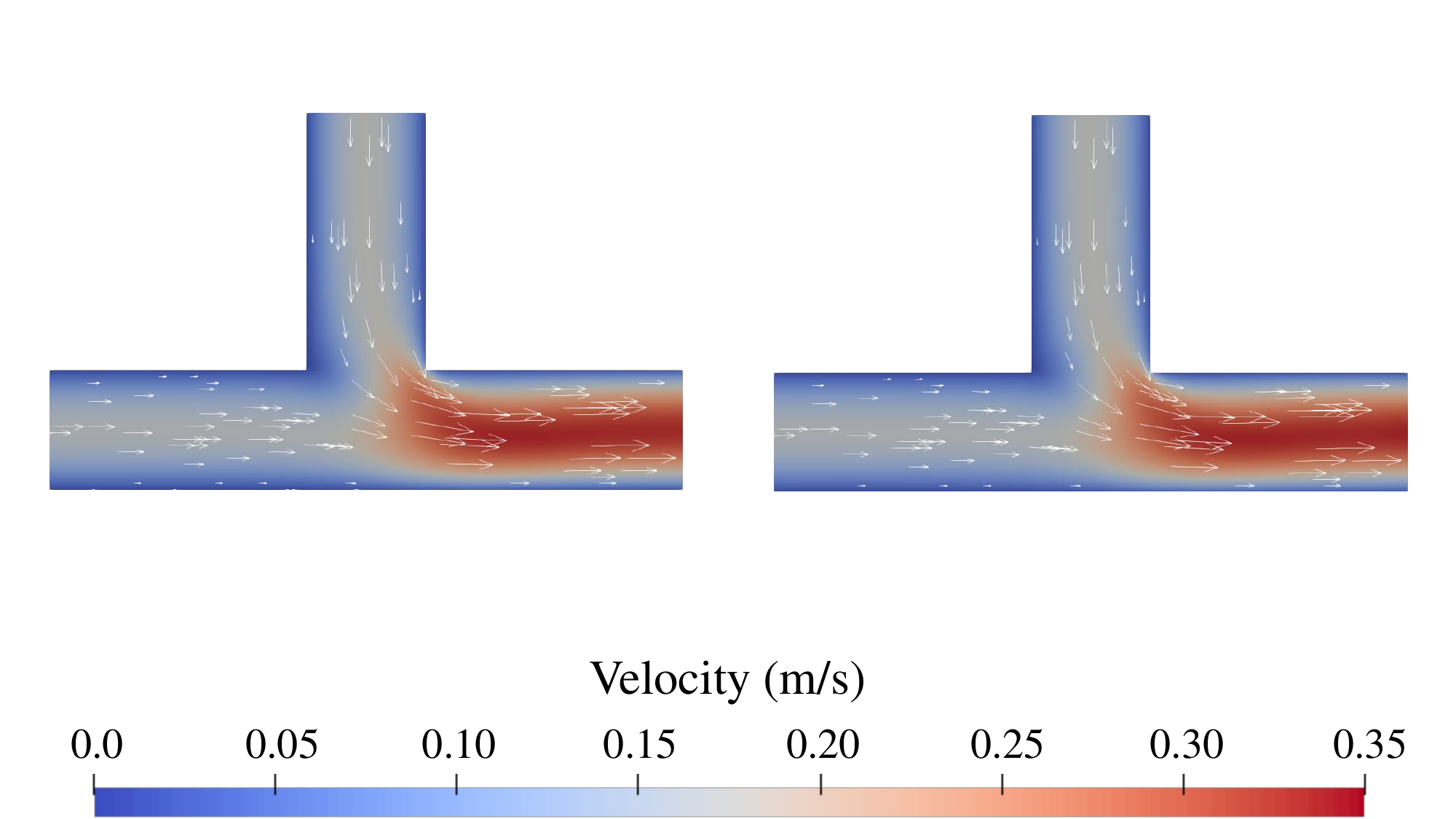}
         \caption{Velocity field of Region \emph{3c}.}
         \label{fig:3a_T2_velocity}
     \end{subfigure}
        \caption{{The} 
 pressure and velocity fields obtained from the CFD simulations (\textbf{left}) and the proposed method (\textbf{right}) for Regions \emph{3a}, \emph{3b}, and \emph{3c} in Network 3 at \(l\text{ is }1\). }
        \label{fig:3a}
\end{figure*}

\begin{figure*}[h!]
     \begin{subfigure}[b]{0.48\textwidth}
         \centering
         \includegraphics[width=1.0\textwidth]{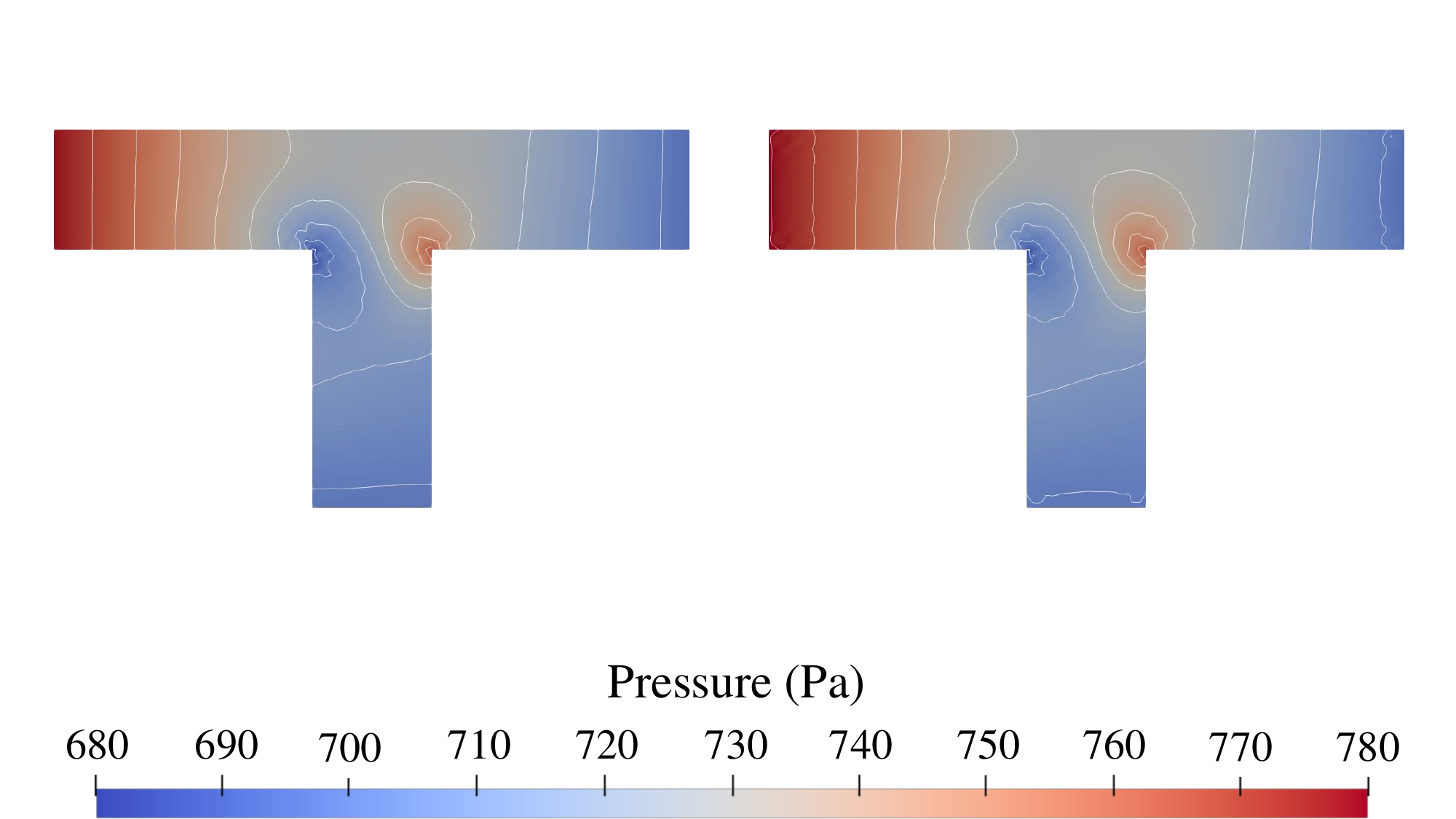}
         \caption{Pressure field of Region \emph{4a}.}
         \label{fig:4a_invT0_pressure}
     \end{subfigure}
     \hfill
     \begin{subfigure}[b]{0.48\textwidth}
         \centering
         \includegraphics[width=1.0\textwidth]{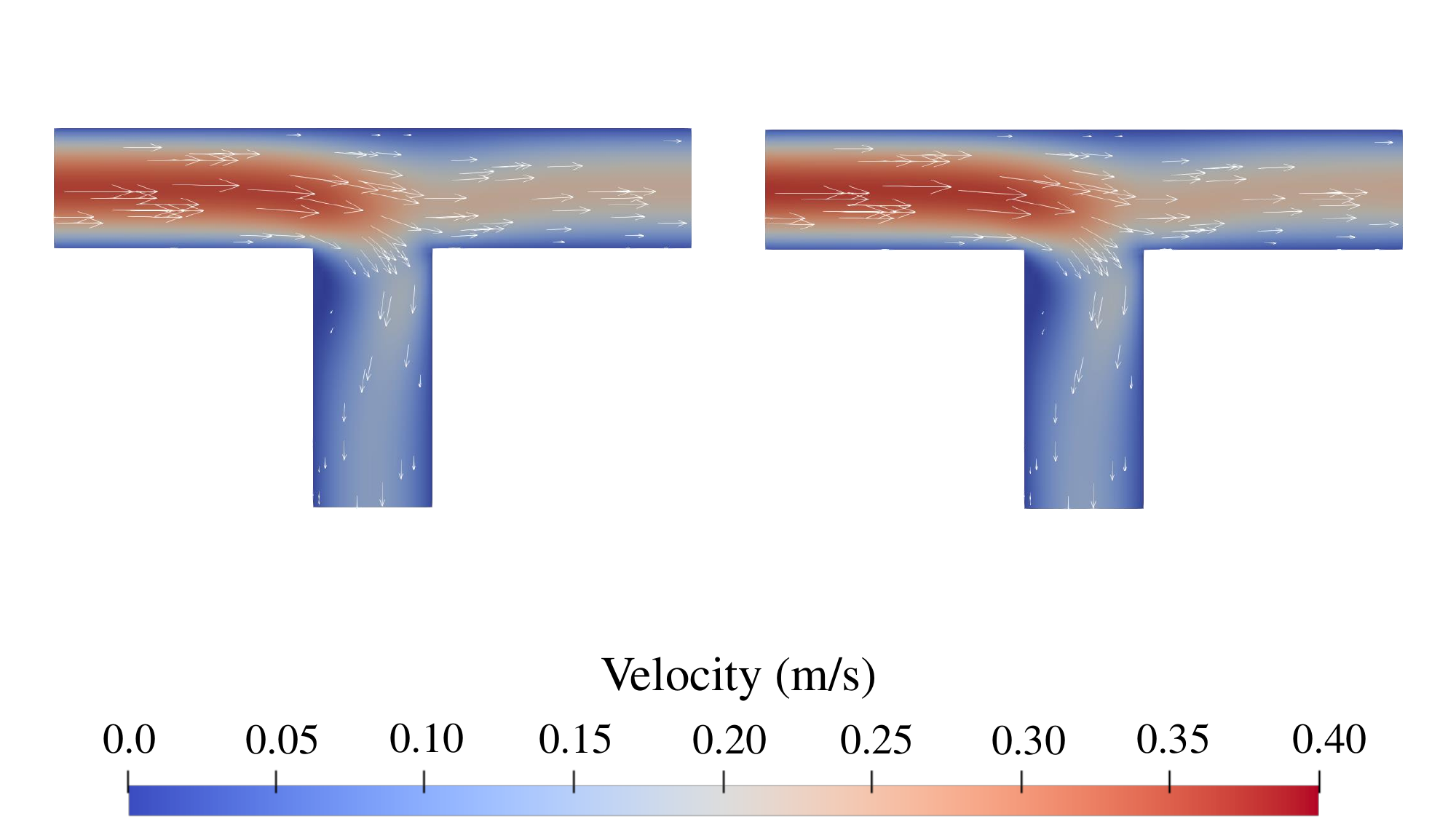}
         \caption{Velocity field of Region \emph{4a}.}
         \label{fig:4a_invT0_velocity}
     \end{subfigure}\\
     \\
    \begin{subfigure}[b]{0.48\textwidth}
         \centering
         \includegraphics[width=1.0\textwidth]{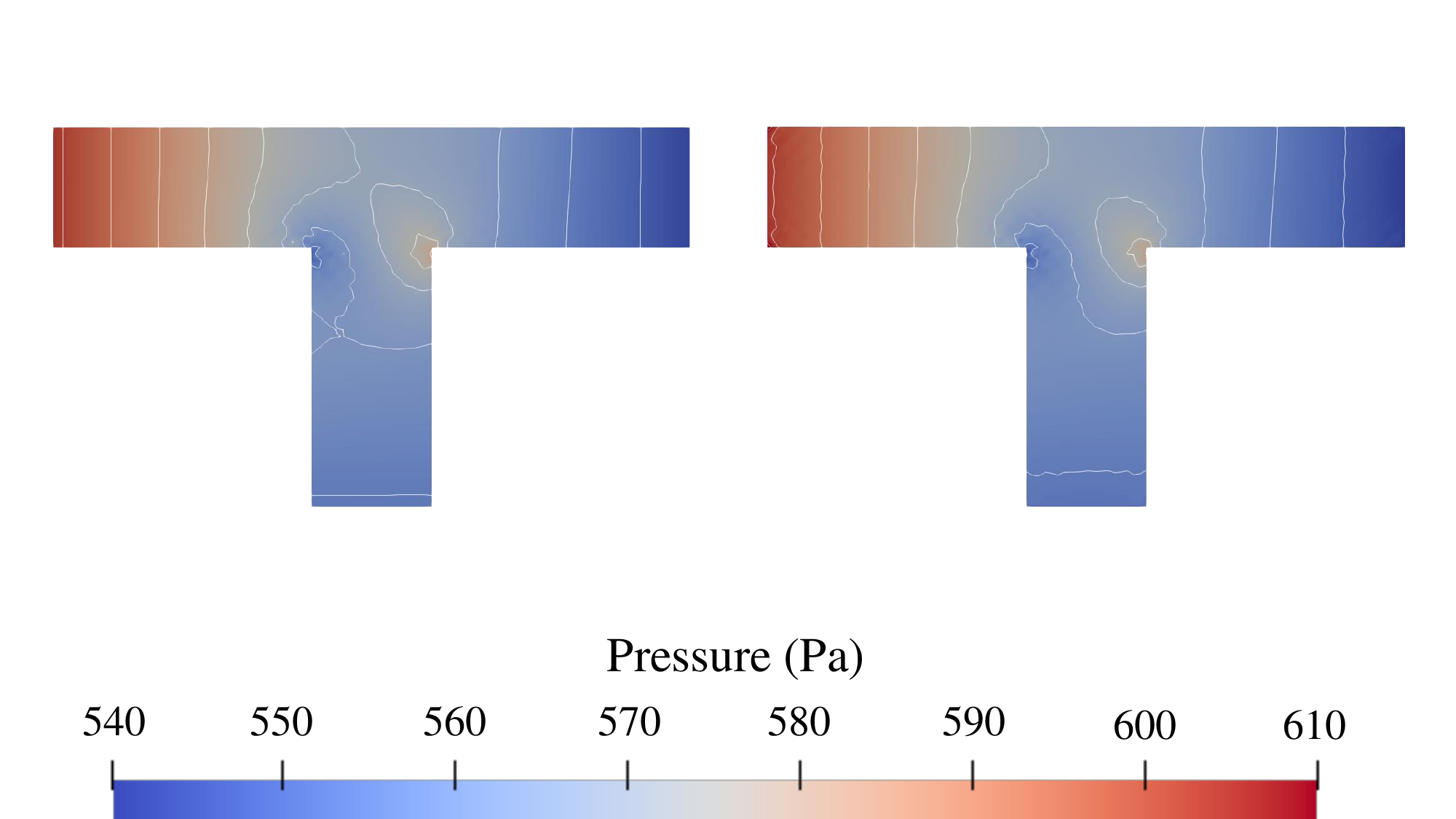}
         \caption{Pressure field of Region \emph{4b}.}
         \label{fig:4a_invT1_pressure}
     \end{subfigure}
     \hfill
     \begin{subfigure}[b]{0.48\textwidth}
         \centering
         \includegraphics[width=1.0\textwidth]{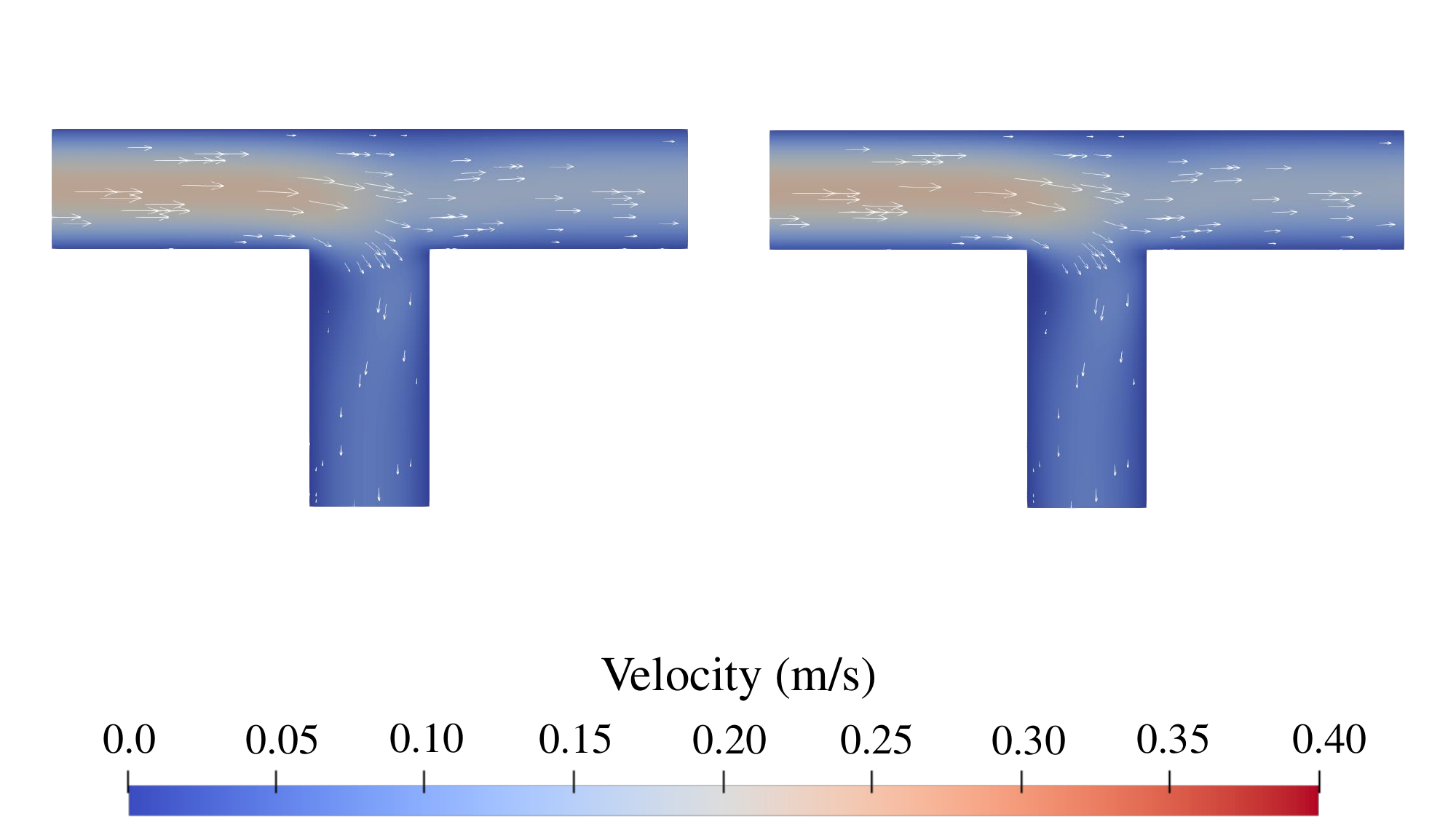}
         \caption{Velocity field of Region \emph{4b}.}
         \label{fig:4a_invT1_velocity}
     \end{subfigure}\\
     \\
    \begin{subfigure}[b]{0.48\textwidth}
         \centering
         \includegraphics[width=1.0\textwidth]{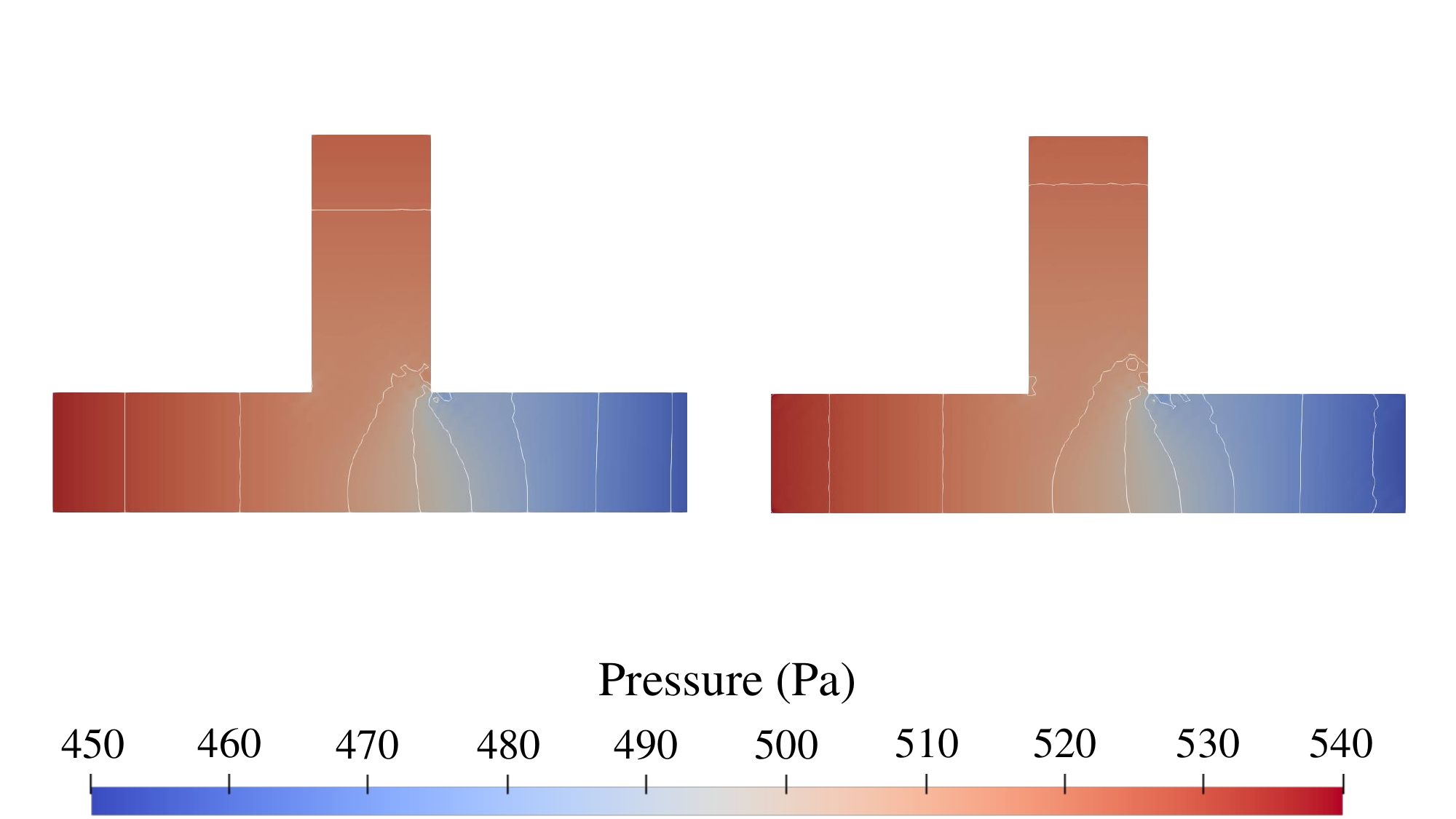}
         \caption{Pressure field of Region \emph{4c}.}
         \label{fig:4a_T2_pressure}
     \end{subfigure}
     \hfill
     \begin{subfigure}[b]{0.48\textwidth}
         \centering
         \includegraphics[width=1.0\textwidth]{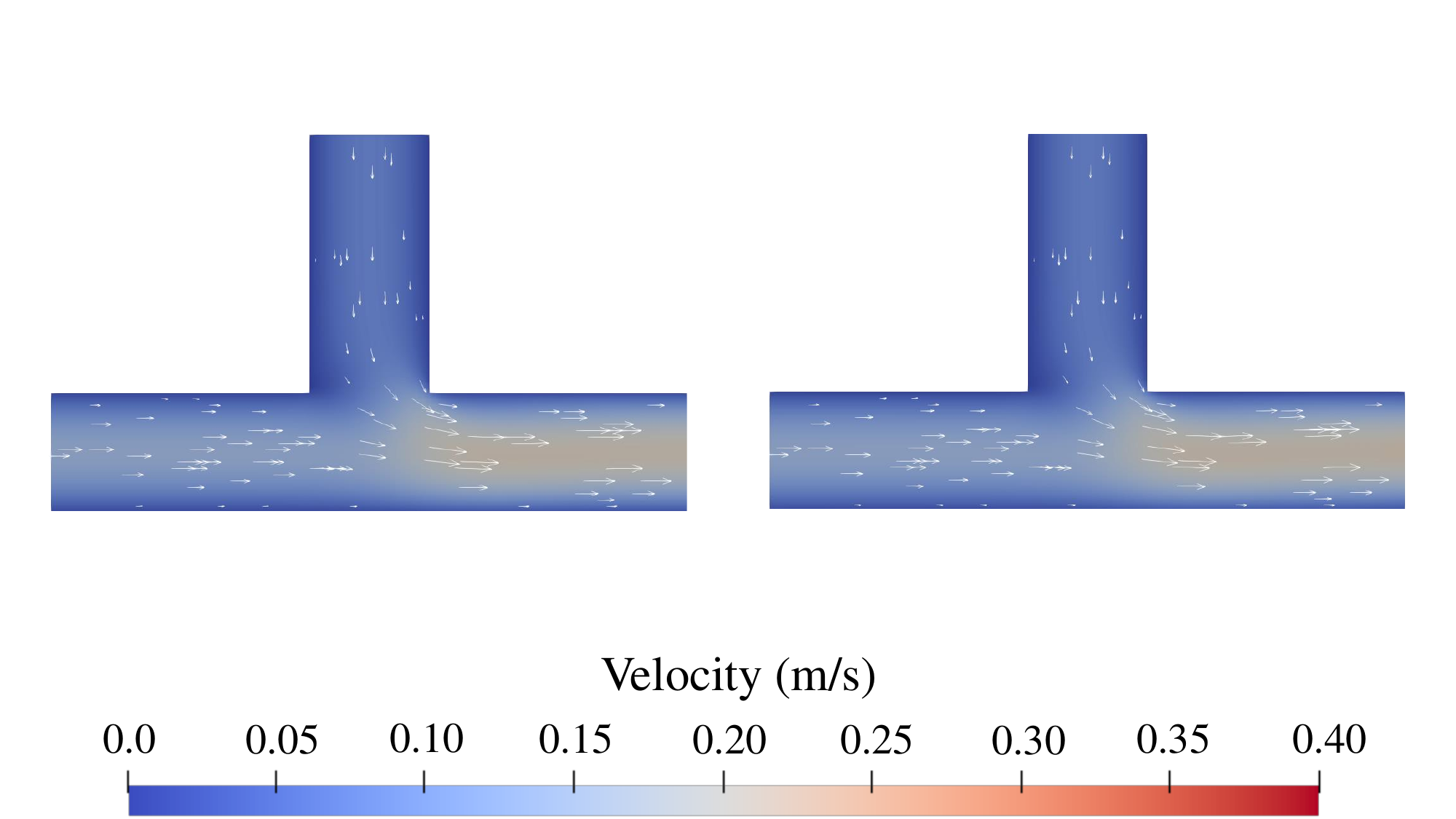}
         \caption{Velocity field of Region \emph{4c}.}
         \label{fig:4a_T2_velocity}
     \end{subfigure}\\
     \\
    \begin{subfigure}[b]{0.48\textwidth}
         \centering
         \includegraphics[width=1.0\textwidth]{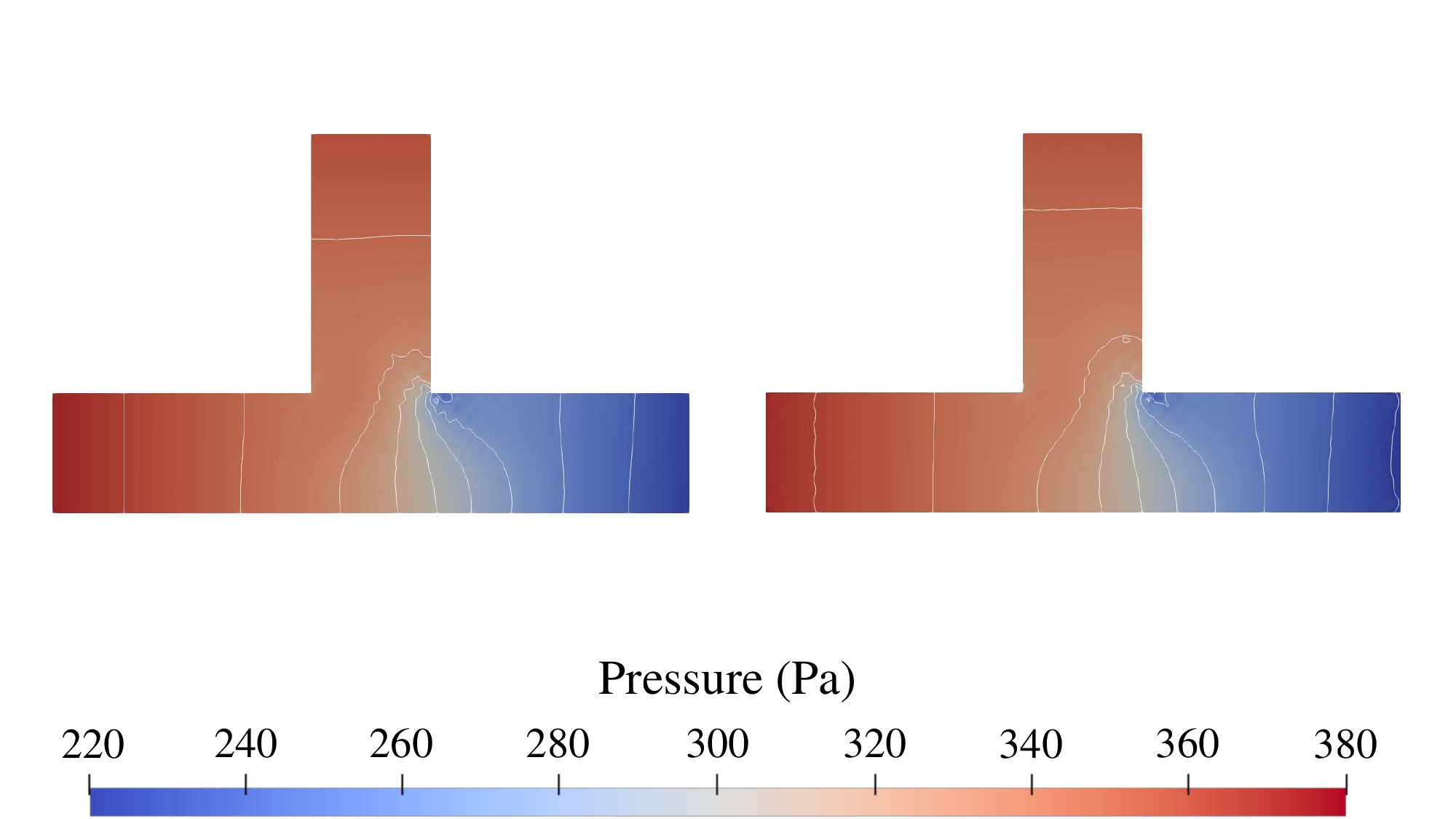}
         \caption{Pressure field of Region \emph{4d}.}
         \label{fig:4a_T3_pressure}
     \end{subfigure}
     \hfill
     \begin{subfigure}[b]{0.48\textwidth}
         \centering
         \includegraphics[width=1.0\textwidth]{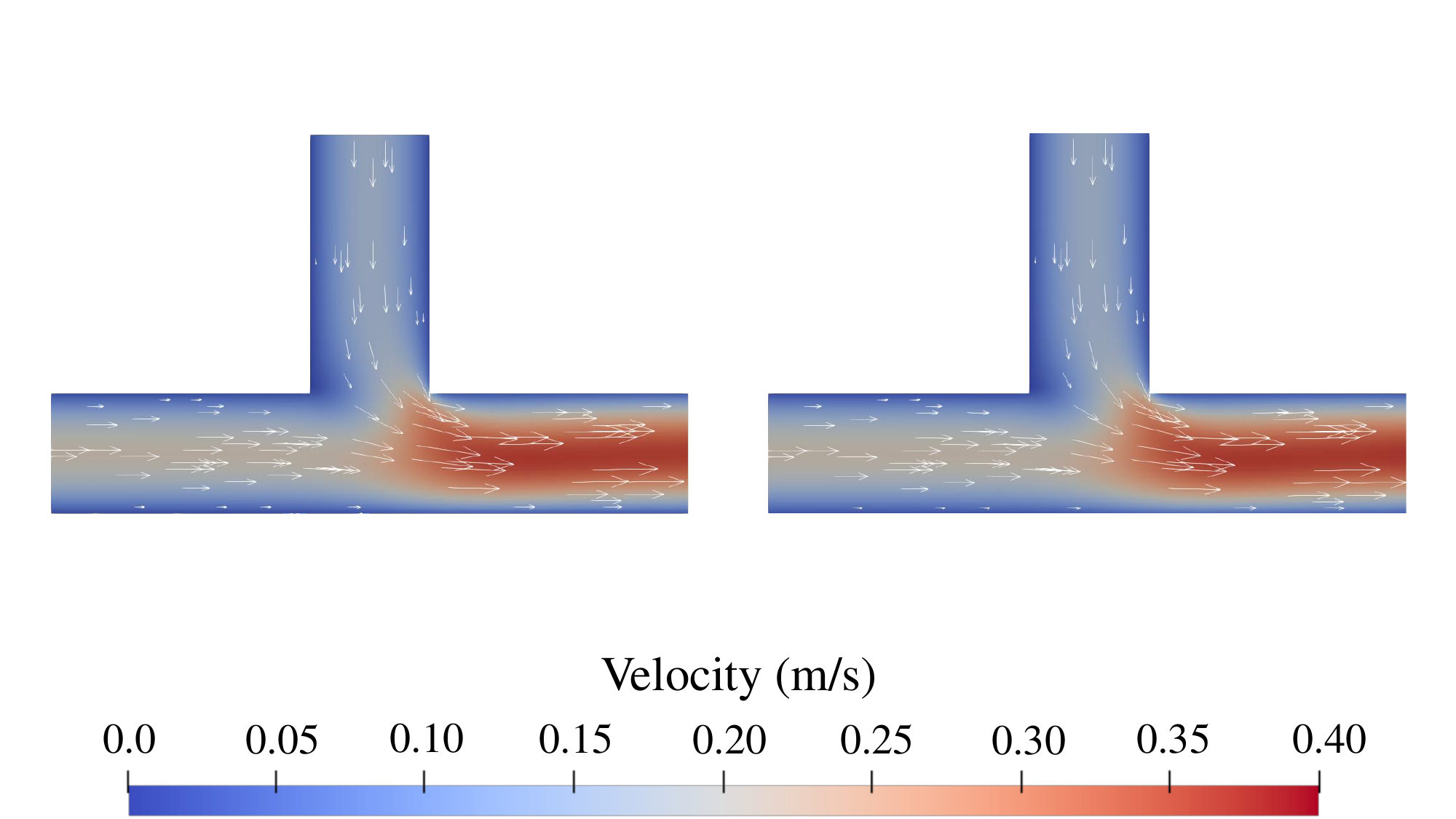}
         \caption{Velocity field of Region \emph{4d}.}
         \label{fig:4a_T3_velocity}
     \end{subfigure}
        \caption{{The} 
 pressure and velocity fields obtained from the CFD simulations (\textbf{left}) and the proposed method (\textbf{right}) for Regions \emph{4a}, \emph{4b}, \emph{4c}, and \emph{4d} in Network 4 at \(l\text{ is }1\). }
        \label{fig:4a}
\end{figure*}

Concerning accuracy, direct comparisons of the pressure and velocity fields obtained with CFD simulations and the proposed method are given in Figures \ref{fig:1_cross0}--\ref{fig:4a}. \cref{fig:1_cross0} shows all the obtained results for Region 1a of Network 1 (\cref{fig:PaperNetworks}a) for all four channel lengths \(l\), i.e., the complete {Network 1}
-column in \cref{tab:Runtimes}. \cref{fig:2a,fig:3a,fig:4a}, respectively, show the obtained results for all $\Omega_\text{low}$-regions in Networks 2, 3, and 4 \cref{fig:PaperNetworks}b--d, at channel length \(l = 1\), i.e., the top row in \cref{tab:Runtimes}. Finally, we list the pressure values and velocity magnitudes obtained by both approaches for all test cases in \cref{tab:Verification,tab:VerificationVel}. These values were taken in the $\Omega_\text{low}$-regions (labeled 1a, 2a, ..., 4c, 4d in \cref{fig:PaperNetworks}) at the measuring points indicated by the black dots in \cref{fig:PaperNetworks}. Since the networks were simulated for four different channel lengths $l$, each measuring point has four pressure values.

\begin{table*}
    \small
	\centering
		\caption{Required runtimes of the original CFD simulations and the proposed method with corresponding speed-ups.}
	\label{tab:Runtimes}
	\def\arraystretch{1}
	\setlength{\tabcolsep}{4pt}
	\makebox[\linewidth]{
		\begin{tabular}{@{\,}l@{\qquad}ccc@{\qquad}ccc@{\qquad}ccc@{\qquad}ccc@{\,}}
            \toprule
			& \multicolumn{3}{c}{\textbf{Network 1}\hspace*{0pt}} & \multicolumn{3}{c}{\textbf{Network 2}} & \multicolumn{3}{c}{\textbf{Network 3}} & \multicolumn{3}{c}{\textbf{Network 4}} \\
			\cmidrule{2-13} 
			& \multicolumn{2}{c}{\shortstack{\textbf{Runtime}\\ {\textbf{[hh:mm:ss]}}}\hspace*{0pt}} & \multirow{1.75}{*}{\textbf{Speed-Up}}
			& \multicolumn{2}{c}{\shortstack{\textbf{Runtime}\\ {\textbf{[hh:mm:ss]}}}\hspace*{0pt}} & \multirow{1.75}{*}{\textbf{Speed-Up}}
			& \multicolumn{2}{c}{\shortstack{\textbf{Runtime}\\ {\textbf{[hh:mm:ss]}}}\hspace*{0pt}} & \multirow{1.75}{*}{\textbf{Speed-Up}}
			& \multicolumn{2}{c}{\shortstack{\textbf{Runtime}\\ {\textbf{[hh:mm:ss]}}}\hspace*{0pt}} & \multirow{1.75}{*}{\textbf{Speed-Up}}\\
			\cmidrule(r{5pt}){2-3} \cmidrule(r{5pt}){5-6} \cmidrule(r{5pt}){8-9} \cmidrule(r{5pt}){11-12}
			\boldmath{$l$} & \textbf{CFD} & \textbf{Proposed} & 
			& \textbf{CFD} & \textbf{Proposed} & 
			& \textbf{CFD} & \textbf{Proposed} &   
			& \textbf{CFD} & \textbf{Proposed} & \\
			\midrule
			1 
			& 03:15:51 & 00:02:04 & 94.7
			& 01:34:08 & 00:09:07 & 10.3
			& 04:11:49 & 00:20:05 & 12.5  
			& 02:21:41 & 00:44:28 & 3.2  \\
			2 
			& 13:23:38 & 00:02:12 & 365.2
			& 06:16:43 & 00:08:55 & 42.2 
			& 11:55:38 & 00:15:41 & 55.0
			& 07:56:19 & 00:59:12 & 8.0 
			\\
			3 
			& 19:48:30 & 00:02:26 & 487.0
			& 14:10:54 & 00:08:44 & 81.9 
			& 28:32:50 & 00:15:57 & 107.3  
			& 16:16:49 & 01:07:49 & 14.4 \\
			4 
			& 52:27:35 & 00:02:58 & 1055.4
			& 21:48:48 & 00:08:13 & 159.3
			& 49:55:21 & 00:15:39 & 191.2
			& 29:03:13 & 01:19:53 & 21.8 \\
			\bottomrule
		\end{tabular}
	}
\end{table*}

\begin{table}
\small
	\centering
		\caption{Obtained pressure values of all test-cases at the measuring points (as denoted in \cref{fig:ExampleNetwork1,fig:ExampleNetwork2,fig:ExampleNetwork3,fig:ExampleNetwork4}) obtained from  the original CFD simulation and the proposed method.}
	\label{tab:Verification}
	\setlength{\tabcolsep}{4pt}
	\setlength{\cmidrulewidth}{\lightrulewidth}
	\begin{tabular}{clcc@{\qquad\qquad}clcc}
        \toprule
		& & \multicolumn{2}{c}{\textbf{Pressure {[Pa]}}\hspace*{30pt}} & & & \multicolumn{2}{c}{\textbf{Pressure {[Pa]}}\hspace*{0pt}} \\
		\cmidrule{3-8} 
		\boldmath{$\Omega_{low}$} & \boldmath{$l$} & \textbf{CFD} & \textbf{Proposed} & \boldmath{$\Omega_{low}$} & \boldmath{$l$} & \textbf{CFD} & \textbf{Proposed} \\
		\midrule
		\multirow{3.75}{*}{1a} & 1 & 710.1 & 712.6 & \multirow{3.75}{*}{3c} & 1 & 296.5 & 294.4 \\
		& 2 & 680.8 & 681.8 &  & 2 & 275.0 & 272.9 \\
		& 3 & 552.7 & 553.9 &  & 3 & 270.7 & 268.7 \\
		& 4 & 670.6 & 671.0 &  & 4 & 269.1 & 267.2 \\
		\cmidrule{1-8}
		\multirow{3.75}{*}{2a} & 1 & 661.8 & 662.3 & \multirow{3.75}{*}{4a} & 1 & 729.1 & 728.9 \\
		& 2 & 643.6 & 643.6 &  & 2 & 712.0 & 711.4 \\
		& 3 & 639.6 & 639.5 &  & 3 & 708.5 & 707.7 \\
		& 4 & 638.2 & 637.9 &  & 4 & 707.3 & 706.3 \\
		\cmidrule{1-8}
		\multirow{3.75}{*}{2b} & 1 & 398.9 & 397.8 & \multirow{3.75}{*}{4b} & 1 & 568.9 & 567.0 \\
		& 2 & 373.4 & 372.2 &  & 2 & 539.6 & 537.8 \\
		& 3 & 368.2 & 367.0 &  & 3 & 533.9 & 532.0 \\
		& 4 & 366.3 & 365.1 &  & 4 & 532.0 & 529.9 \\
		\cmidrule{1-8}
		\multirow{3.75}{*}{3a} & 1 & 748.1 & 748.0 & \multirow{3.75}{*}{4c} & 1 & 508.5 & 506.7 \\
		& 2 & 736.8 & 736.5 &  & 2 & 480.7 & 478.6 \\
		& 3 & 734.7 & 734.1 &  & 3 & 475.3 & 473.1 \\
		& 4 & 734.0 & 733.3 &  & 4 & 473.3 & 471.0 \\
		\cmidrule{1-8}
		\multirow{3.75}{*}{3b} & 1 & 565.3 & 563.0 & \multirow{3.75}{*}{4d} & 1 & 325.9 & 323.8 \\
		& 2 & 541.7 & 539.8 &  & 2 & 302.7 & 300.4 \\
		& 3 & 537.0 & 535.3 &  & 3 & 298.2 & 296.0 \\
		& 4 & 535.5 & 533.7 &  & 4 & 296.5 & 294.4 \\
		\bottomrule
	\end{tabular}
\end{table}

\begin{table}
\small
	\centering
		\caption{{Obtained} velocity magnitudes of all test-cases at the measuring points (as denoted in \cref{fig:PaperNetworks}a--d) obtained from  the original CFD simulation and the proposed method.}
	\label{tab:VerificationVel}
	\setlength{\tabcolsep}{4pt}
	\setlength{\cmidrulewidth}{\lightrulewidth}
	\begin{tabular}{clcc@{\qquad\qquad}clcc}
        \toprule
		& & \multicolumn{2}{c}{\textbf{Velocity {[mm/s]}}\hspace*{30pt}} & & & \multicolumn{2}{c}{\textbf{\textbf{Velocity {[mm/s]}}}\hspace*{0pt}} \\
		\cmidrule{3-8} 
		\boldmath{$\Omega_{low}$} & \boldmath{$l$} & \textbf{CFD} & \textbf{Proposed} & \boldmath{$\Omega_{low}$} & \boldmath{$l$} & \textbf{CFD} & \textbf{Proposed} \\
		\midrule
		\multirow{3.75}{*}{1a} & 1 & 363.3 & 369.0 & \multirow{3.75}{*}{3c} & 1 & 192.3 & 192.4 \\
		& 2 & 203.6 & 207.8 &  & 2 & 102.2 & 99.3 \\
		& 3 & 152.7 & 155.3 &  & 3 & 69.9 & 70.2 \\
		& 4 & 106.8 & 109.0 &  & 4 & 53.2 & {53.5} \\ 
		\cmidrule{1-8} 
		\multirow{3.75}{*}{2a} & 1 & 378.6 & 384.9 & \multirow{3.75}{*}{4a} & 1 & 280.52 & 285.8 \\
		& 2 & 183.0 & 186.8 &  & 2 & 136.2 & 139.5 \\
		& 3 & 119.6 & 122.3 &  & 3 & 89.3 & 91.6 \\
		& 4 & 88.6 & 90.6 &  & 4 & 66.3 & {68.1} \\ 
		\cmidrule{1-8} 
		\multirow{3.75}{*}{2b} & 1 & 344.0 & 346.7 & \multirow{3.75}{*}{4b} & 1 & 178.2 & 180.3 \\
		& 2 & 169.3 & 171.1 &  & 2 & 84.0 & 85.3 \\
		& 3 & 112.4 & 113.6 &  & 3 & 54.9 & 55.8 \\
		& 4 & 84.2 & 85.2 &  & 4 & 40.8 & {41.5} \\ 
		\cmidrule{1-8} 
		\multirow{3.75}{*}{3a} & 1 & 274.8 & 279.8 & \multirow{3.75}{*}{4c} & 1 & 143.6 & 144.7 \\
		& 2 & 132.5 & 130.8 &  & 2 & 76.3 & 77.2 \\
		& 3 & 86.7 & 88.1 &  & 3 & 51.6 & 52.3 \\
		& 4 & 64.3 & 66.0 &  & 4 & 38.9 & {39.5} \\ 
		\cmidrule{1-8} 
		\multirow{3.75}{*}{3b} & 1 & 181.5 & 183.2 & \multirow{3.75}{*}{4d} & 1 & 231.0 & 231.2 \\
		& 2 & 93.0 & 91.1 &  & 2 & 120.9 & 121.4 \\
		& 3 & 62.3 & 63.3 &  & 3 & 81.8 & 82.3 \\
		& 4 & 46.8 & 47.5 &  & 4 & 62.0 & 62.3 \\
		\bottomrule
	\end{tabular}
\end{table}

\clearpage

\clearpage

The results clearly confirm the improvement and benefit of the proposed acceleration method. 
First, we can see from the numbers summarized in \cref{tab:Verification,tab:VerificationVel} that both the original CFD simulation, as well as the proposed method, more or less provide the same results. That is, using higher levels of abstractions for $\Omega_\text{high}$-regions does not significantly affect the simulation results. At the same time, this enables impressive speed-ups. In fact, the numbers summarized in \cref{tab:Runtimes} do not only show that the proposed method \emph{{always}} generates the results faster than the original CFD method, but also that it constantly does so by several factors---in many cases even by up to three orders of magnitudes. 

\subsection{Discussion}
The obtained results, as summarized above, clearly show the benefits of the proposed method. On top of that, they also provide further, more detailed insights, as well as implications and ideas for further extensions. These are discussed in the following.

Firstly, with respect to the accuracy, the results presented in \cref{fig:1_cross0,fig:2a,fig:3a,fig:4a} and \mbox{Tables}~\ref{tab:Verification}~and~\ref{tab:VerificationVel} show a strong alignment between the proposed method and the corresponding CFD simulations. From \cref{fig:1_cross0}, it can be concluded that there seems to be a slight improvement in the pressure field as the channel length \(l\) increases, but this is negligible. The results in \cref{fig:2a,fig:3a,fig:4a} show that the method is applicable for problems with multiple disconnected $\Omega_\text{high}$-regions, with a negligible increase in inaccuracy for $\Omega_\text{high}$-regions that are located between others, such as 3b, 4b and 4c. Following from the pressure contour lines obtained from the CFD approach, which generally appear straight towards \textGamma{}, the location of \textGamma{} can be said to be sufficiently far away from the crossings and junctions and did not strongly influence the performance of the proposed method. The pressure contour lines obtained from the proposed approach, however, consistently show a curvature near \textGamma{}, which can be attributed to numerical intricacies of the underlying implementation of the boundary conditions. This slight inaccuracy can also {explain} 
the propagated error to Regions 3b, 4b, and 4c. 

The exploitation of higher levels of abstraction is possible due to the availability of reduced order modeling methods for microfluidic flow. In this work, the Hagen--Poiseuille law was used to represent the fluid flow through straight two-dimensional channels. Using a different method of high abstraction, that models flow through three-dimensional channels, such as presented in \cite{Cornish1928}, would allow the proposed method to be extended to three dimensional problems. With this added dimension and, therefore, even worse computational complexity of CFD simulations, the speed-up of the proposed method can be expected to be of even higher orders of magnitude. Provided that methods of high abstraction exist, the proposed method can even be extended to include other physical phenomena, such as diffusion, heat dissipation, or droplets. For straight channels, diffusion and heat dissipation can simply be modeled as a \mbox{time-dependent} transport across the channel width, and high abstraction simulation approaches for droplets are performed by adding the droplet hydraulic resistance to the channel \cite{Vanapalli2009}. Further work is required to include additional physical phenomena in the proposed method.

\section{Conclusion}
\label{sec:Conclusion}
In this work, we proposed an accelerated CFD simulation method for microfluidic devices. The core idea was to utilize higher levels of abstraction whenever possible to improve the simulation runtime while maintaining the fidelity. We developed and implemented a prototype of the resulting simulation approach, using continuous channel-based microfluidic devices as a representative platform. Results obtained from corresponding case studies confirmed the promises of the proposed approach: using higher levels of abstractions for the simulation did not significantly affect the fidelity of the simulation results, but allowed for substantial \mbox{speed-ups} of up to three orders of magnitude. Based on this premise, and with the inclusion of other physical phenomena such as diffusion, heat dissipation, or multiphase flow (e.g., droplets), similar accelerations can be expected for further microfluidic platforms, which is left for future work. Overall, this provides the foundation for more research toward exploiting higher levels of abstraction for simulating microfluidic devices in future work.

\section*{Author Contributions}
Conceptualization, M.T. and R.W.; methodology, M.T.; software, M.T.; writing—original draft preparation, M.T.; writing—review and editing, R.W. All authors have read and agreed to the published version of the manuscript.

\section*{Conflicts of Interest}
The authors declare no conflict of interest. The funders had no role in the design of the study; in the collection, analyses, or interpretation of data; in the writing of the manuscript, or in the decision to publish the results.

\section*{Acknowledgments}
This work has partially been supported by the FFG project AUTOMATE (project number: 890068) as well as by BMK, BMDW, and the State of Upper Austria in the frame of the COMET Programme managed by FFG.

\bibliographystyle{IEEEtran}
{%
\bibliography{%
lit_header,%
lit_rail}%
}

\end{document}